\documentclass[12pt]{article}

\usepackage[T1]{fontenc} % Use 8-bit encoding that has 256 glyphs
\usepackage{lmodern}
\usepackage{microtype} % Slightly tweak font spacing for aesthetics

\usepackage[dvipsnames]{xcolor}
\definecolor{dark-gray}{gray}{0.20}
\definecolor{gray}{gray}{0.30}
\definecolor{light-gray}{gray}{0.80}
\definecolor{dark-red}{rgb}{0.7,0,0}
\definecolor{dark-green}{rgb}{0.1,0.4,0}
\definecolor{dark-blue}{rgb}{0.3,0.3,0.7}
\definecolor{light-blue}{rgb}{0.8,0.8,1}

\usepackage{cite}    % [nosort] to keep the chronological order
\usepackage[pdfencoding=auto,pagebackref]{hyperref}
\usepackage{hyperref}
\hypersetup{
	colorlinks=true,
	linkcolor=dark-blue,
	citecolor=dark-red,
	urlcolor=dark-green,
	linktoc=page
}

\usepackage{enumerate}
\usepackage{graphicx}
\usepackage[percent]{overpic}
\usepackage{booktabs}
\usepackage{bbm}
\usepackage[margin=15pt,small]{caption}
\usepackage{subcaption}
\usepackage{soul}

\usepackage[percent]{overpic}
\graphicspath{{./Figures/}}

\usepackage{amsmath,amssymb,slashed,amsthm,mathrsfs}
\numberwithin{equation}{section}
\usepackage{slashed}
\usepackage{mathabx}
\usepackage{bm}

\usepackage{braket}
\usepackage{clipboard}
\usepackage{tikz}
\usetikzlibrary{calc,fadings}
\pgfdeclarefading{fad}
{\tikz
	\fill[bottom color=transparent!0,top color=transparent!100] (0,0) rectangle (50bp,50bp);}

\newcommand{\q}[1]{``#1''}
\numberwithin{equation}{section}

\newcommand{\be}{\begin{equation}} \newcommand{\ee}{\end{equation}}
\newcommand{\bea}{\begin{equation} \begin{aligned}} \newcommand{\eea}{\end{aligned} \end{equation}}

\newcommand{\cC}{\mathcal{C}}

\newcommand{\cF}{\mathcal{F}}

\newcommand{\cJ}{\mathcal{J}}

\newcommand{\cN}{\mathcal{N}}
\newcommand{\cO}{\mathcal{O}}

\newcommand{\cQ}{\mathcal{Q}}
\newcommand{\cR}{\mathcal{R}}

\newcommand{\bZ}{\mathbb{Z}}

\def\repa{\raise4pt\hbox{$\square$}\mkern-14mu\raise-4pt\hbox{$\square$}}
\def\repab{\overline{\raise4pt\hbox{$\square$}\mkern-14mu\raise-4pt\hbox{$\square$}\mkern-1mu}}

\renewcommand{\baselinestretch}{1.2}

\begin{document}
	
	\begin{center}
		{\huge \bf $\mathcal{N}=2$ Super Yang-Mills in AdS$_4$\\[-0.25em] and \\[0.25em]
        $F_{\text{AdS}}$-maximization}			
	\end{center}

\vspace{1 cm}

\begin{center}
Davide Bason$^a$, Christian Copetti$^b$, Lorenzo Di Pietro$^{c, d}$ and Ziming Ji$^{d,e,f}$

\vspace{1 cm}
{

${}^a\!\!$
{\em  Yau Mathematical Sciences Center, Tsinghua University, \\ Jingzhai, Haidan District, Beijing, 100084, China}

\vspace{.3cm}

${}^b\!\!$
{\em Mathematical Institute, University of Oxford, \\
	Andrew Wiles Building,  Woodstock Road, Oxford, OX2 6GG, UK}

\vspace{.3cm}

${}^c\!\!$
{\em  Dipartimento di Fisica, Universit\`a di Trieste, \\ Strada Costiera 11, I-34151 Trieste, Italy}

\vspace{.3cm}

${}^d\!\!$
{\em INFN, Sezione di Trieste, Via Valerio 2, I-34127 Trieste, Italy}

\vspace{.3cm}

${}^e\!\!$
{\em SISSA, Via Bonomea 265, 34136 Trieste, Italy }

\vspace{.3cm}

${}^f\!\!$
{\em Department of Physics, Northeastern University, Boston, MA, 02115, USA }

 }
\end{center}

\vspace{1cm}

\begin{center}{\bf Abstract} \end{center}
\vspace{2 mm}
\begin{quote}
We investigate the dynamics of four-dimensional $\mathcal{N}=2$ $SU(2)$ super Yang--Mills theory on an AdS background. We propose that the boundary conditions that preserve the AdS super-isometries are determined by maximizing the real part of the AdS partition function $F_{\text{AdS}}=-\log Z_{\text{AdS}}$. At weak coupling $\Lambda L \ll 1$ the maximization singles out the Dirichlet boundary condition with an $SU(2)$ boundary global symmetry, corresponding to the classical vacuum at the origin of the Coulomb branch with fully un-higgsed gauge group. We find that for $\Lambda L \sim \mathcal{O}(1)$ new boundary conditions are favored, with gauge-group higgsed down to $U(1)$, matching the expectation from the flat space limit.  We use supersymmetric localization to compute $Z_{\text{AdS}}$ nonperturbatively. We further provide evidence for a relation between $F_{\text{AdS}}$ and the $\cN=2$ prepotential in AdS background.
\end{quote}

\newpage

\renewcommand{\baselinestretch}{1.1}

\small 

\tableofcontents

\normalsize
\newpage

\renewcommand{\baselinestretch}{1.2}

\section{Introduction}\label{sintro}
Using Anti-de Sitter space as a playground to study asymptotically-free QFTs --and in particular gauge theories-- has a long history, dating back to \cite{Callan:1989em} and, more recently, to \cite{Aharony:2012jf}. At the core of this program is the realization that the bulk isometry group $SO(1, d+1)$ of AdS$_{d+1}$ acts as the conformal group on the asymptotic boundary degrees of freedom. This is a powerful tool to constrain the asymptotic observables, which can be eventually connected to the flat space ones by taking the limit of large radius \cite{Paulos:2016fap, Komatsu:2020sag, vanRees:2022zmr}.

In \cite{Aharony:2012jf} it was proposed that intrinsically strong coupling bulk processes, such as dynamical mass generation, can be equivalently described by a boundary transition between different boundary conditions, using the language and the tools that are familiar from the study of conformal field theories. 
For a confining gauge theory in AdS, this is a consequence of the fact that, subject to the Dirichlet boundary condition, the bulk gauge fields remain massless and there are colored asymptotic states. In order to match the flat-space expectation, a transition must take place as we enter the strong coupling region $\Lambda L \sim \mathcal{O}(1)$, where $\Lambda$ is the strong coupling scale and $L$ the AdS length, whereby the Dirichlet boundary condition stops being an available boundary condition compatible with unitarity and AdS isometries. 

There are various proposals for the precise nature of the transition \cite{Aharony:2012jf}. A first possibility is that the boundary conserved current $J^\mu$ associated to the Dirichlet boundary condition is continuously lifted by a multiplet recombination, corresponding to a continuous transition to a Higgs phase in the bulk.\footnote{Note that even in the presence of such a continuous recombination, the Dirichlet boundary condition would keep existing in addition to the Higgsed boundary condition after the recombination, and its fate in the strong coupling limit would still remain to be determined.} On the other hand the recent studies \cite{Copetti:2023sya,Ciccone:2024guw} suggest that a boundary operator, singlet of the global symmetries and irrelevant at weak coupling, reaches marginality at a critical value $\Lambda L\sim \mathcal{O}(1)$ and causes the Dirichlet boundary condition to disappear by merger and annihilation \cite{Kaplan:2009kr, Gorbenko:2018ncu}. This mechanism has been so far shown to hold in 2d sigma models and the Gross-Neveau model in the large $N$ limit, and it is favored by the perturbative results for scaling dimensions in pure 4d Yang-Mills theory. The mechanism by which the appearance of such a marginal operator generically leads to breaking AdS isometries or unitarity is explained in \cite{Lauria:2023uca}.

The quantitative studies of these boundary transitions so far used the large-$N$ expansion \cite{Copetti:2023sya} or indications from perturbation theory \cite{Ciccone:2024guw}. Given that the interesting dynamics happens at strong coupling, it is precious to have examples in which exact nonperturbative techniques are available. Motivated by this, in this work we extend the study of QFT in AdS to the case of supersymmetric gauge theories in four dimensions, in which supersymmetry allows to obtain exact results. The minimal amount that allows us to use localization techniques is $\mathcal{N}=2$ extended supersymmetry.

We will concentrate on the simplest example, namely the  $\cN=2$ pure $SU(2)$ Super Yang-Mills (SYM) theory in four dimensions. This theory does not confine in flat space, nonetheless its behavior at strong coupling defeats the semiclassical intuition and it allows us to pose a sharp question about the behavior of its AdS boundary conditions. Let us recall some of the salient features of the flat space behavior that we will use in our analysis:  in flat space the theory has a Coulomb branch of supersymmetric vacua, parametrized by the VEV of the scalar $\phi$ in the $\cN=2$ vector multiplet\footnote{The minus sign is due to our conventions on the scalar field $\phi$, that we spell out in section \ref{s4dAdSsusy}.}
\be
u = -\left\langle\text{Tr}\left[\phi^2 \right]\right\rangle \, .
\ee
Classically, the origin $u=0$ of the Coulomb branch has a full unbroken $SU(2)$ gauge symmetry, while in every other point the gauge symmetry is broken by an adjoint Higgsing down to $U(1)$, leading to a Coulomb phase at low energy. However, as pointed out by Seiberg and Witten \cite{Seiberg:1994rs}, at the quantum level even the origin of the Coulomb branch $u=0$  breaks the non-abelian gauge symmetry down to $U(1)$, and there is no point with full $SU(2)$ gauge symmetry, contrary to the classical expectation.

When we study the theory with $u=0$ in AdS, the Dirichlet boundary condition is certainly valid in the perturbative regime $\Lambda L \ll 1$,  and it gives an $SU(2)$ global symmetry on the boundary. Similarly to the case of non-supersymmetric Yang-Mills, compatibility with the flat space behavior requires that there should be a strong coupling transition for $\Lambda L \sim \mathcal{O}(1)$ to a new boundary condition. Differently from the case of non-supersymmetric Yang-Mills, this is not expected to be a confining boundary condition, but rather a Higgsed boundary condition with lower $U(1)$ global symmetry, giving the Coulomb phase that is realized in flat space.

Our main result is that we will be able to give strong evidence in support of these expectations, using nonperturbative observables, in particular the supersymmetric partition function that can be computed using localization. While we are primarily motivated by the specific example of $SU(2)$ SYM, along the way we uncover a number of general properties of $\mathcal{N}=2$ field theories in AdS background.

To explain the results in more detail, let us step back to the description of the Coulomb branch in flat space, and its AdS counterpart. A convenient redundant variable to parametrize it in flat space is the VEV of the complex scalar in the vector multiplet $\phi = - \tfrac{i}{2} a \tau_3$. Here $\tau_3$ denotes the unbroken Cartan generator, while $\tau_{\pm}$ denote the broken ones. At the classical level $u = \frac{a^2}{2}$ but this relation gets modified quantum-mechanically. The low energy effective theory in the Coulomb phase is completely encoded in a single holomorphic function of $a \in \mathbb{C}$, the prepotential $\mathcal{F}(a)$. 

In AdS$_4$, the boundary modes of the bulk vector multiplet can be organized in two multiplets $\mathcal{J}_D^{(0)}$ and $\mathcal{J}_N^{(0)}$ of conserved currents, that close under a massive 3d $\mathcal{N}=2$ supersymmetry acting on the boundary $S^3$. See eq.s \eqref{eJD&Ncurrentasymp} for the detailed expressions in terms of the bulk fields. The Dirichlet boundary condition corresponds to fixing the multiplet $\mathcal{J}_D^{(0)}$ to vanish, while the fluctuating multiplet $\mathcal{J}_N^{(0)}$ gives the conserved current of the $SU(2)$ boundary global symmetry. Turning on a VEV $a\in \mathbb{R}$ in AdS corresponds to modifying the Dirichlet boundary condition, by giving a nonzero supersymmetric value to the bottom component component of $\mathcal{J}_D^{(0)}$
\be \label{eq: DirichletAdSintro}
\cJ_{\text{D}}^{(0)}= \left( -a\tau_3, \, 0 , \, 0 , \, 0 , \, 0     \right) \, .
\ee
This modification turns on a real mass deformation for the global symmetry with current $\mathrm{Tr}[\tau_3\mathcal{J}_N^{(0)}]$. The deformation breaks the AdS isometries from the 3d $\mathcal{N}=2$ superconformal algebra to a massive subalgebra. This is due to the fact that the moduli space of the flat space theory is completely lifted by the bulk curvature, at least at the classical level. Naively, one might expect that setting $a = 0$ is sufficient to ensure that the boundary condition satisfies the full AdS (super-)isometries, and that this is the only value corresponding to the origin of the Coulomb branch $u=0$ that we want to study. However, preserving the full $\mathcal{N}=2$ supersymmetry in AdS is more subtle than this, because it requires the addition of suitable boundary terms to ensure the invariance of the action. To write down these boundary terms, it is necessary to introduce a cutoff surface in the radial direction. As a result, only a massive 3d $\mathcal{N}=2$ supersymmetry algebra can be made manifest, the same subalgebra that is preserved by the massive deformation $a$. This opens the question: how do we select AdS boundary conditions that preserves the full bulk supersymmetry, acting as a 3d $\mathcal{N}=2$ superconformal algebra on the boundary? Moreover, in the above discussion we only found an AdS counterpart for a real deformation $a\in\mathbb{R}$, and another natural question is: which AdS quantity corresponds to the imaginary part of $a$?

We find that the answers to these two questions are related. One can define an AdS$_4$ partition function for the theory with modified Dirichlet boundary condition, and this is naturally a function of the parameter $a$
\begin{equation}
F_{\text{AdS}}(a) = - \log Z_{\text{AdS}}~~~\text{with boundary condition \eqref{eq: DirichletAdSintro}.}
\end{equation}
We propose to interpret $F_{\text{AdS}}(a)$ as a holomorphic function of the complex combination $a= m-i\delta$. The real part $m$ is the real mass deformation mentioned above. The imaginary part $\delta$ corresponds to a mixing between the $U(1)_\mathcal{R}$ symmetry and the $U(1)$ subgroup of $SU(2)$ preserved by $a\neq 0$, namely
\begin{equation}
\mathcal{R} \to \mathcal{R} + \delta q~,
\end{equation}
where $\mathcal{R}$ and $q$ are the charges of the $R$-symmetry and of the $U(1)$. We further propose that the boundary conditions that preserve the full 3d $\mathcal{N}=2$ superconformal symmetry are determined by the values $\delta^*$ that maximize 
\begin{equation}
\mathrm{Re}[F_\text{AdS}]  = - \log |Z_{\text{AdS}}|~,
\end{equation}
namely we have
\begin{align}
\begin{split}
\partial_\delta \,\mathrm{Re}[F_\text{AdS}](m-i\delta^*) & = 0~,\\
\partial_\delta^2 \, \mathrm{Re}[F_\text{AdS}](m-i\delta^*) & \leq 0~.
\end{split}
\end{align}
The values of $\delta^*$ determined in this way fix the $U(1)_\mathcal{R}$ superconformal $R$-symmetry to be $\mathcal{R}_\text{SCFT} = \mathcal{R} + \delta^* q$. Note that in a maximum at some $\delta^*\neq 0$ the superconformal $U(1)_\mathcal{R}$ symmetry mixes with a $U(1)$ subgroup of $SU(2)$. This is only possible because the $SU(2)$ is broken to $U(1)$ in any such boundary condition, as can be explicitly verified by looking at the divergence of the current
\be
\partial_\mu J^\mu_\pm = \pm \cO_3 \, ,
\ee
while $J_3^\mu$ remains conserved. This is the manifestation of the adjoint Higgsing at the boundary of AdS.

We call this procedure ``$F_{\text{AdS}}$-maximization''. We conjecture it to hold in general for $\mathcal{N}=2$ supersymmetric field theories in AdS. It is mainly motivated by the analogy to other setups with the same amount of supersymmetry, where one is faced with a similar problem of determining the value of the superconformal $U(1)_{\mathcal{R}}$ charge: 3d $\mathcal{N}=2$ supersymmetric theories on $S^3$ \cite{Jafferis:2010un, Closset:2012vg} (to which our proposal is related by the gravitational decoupling limit $M_{\text{Pl}}\to\infty$ for 3d theories with holographic duals), and 4d $\mathcal{N}=2$ superconformal theories on the hemisphere $HS^4$ \cite{Gaiotto:2014gha,Bason:2023bin} (to which our proposal is related in the special case of superconformal theories in AdS$_4$). 

With the help of $F_{\text{AdS}}$-maximization, the strategy to study the transition between boundary conditions for $SU(2)$ SYM becomes clear. First, we set $m=0$ to avoid introducing explicitly a breaking of the $SU(2)$ symmetry down to $U(1)$. Second, we study the maximization of $\mathrm{Re}[F_\text{AdS}](-i\delta)$ as a function of $\delta$, for various values of the coupling constant $\Lambda L$. This is made possible by the fact that the partition function can be computed exactly using localization. Deriving the expression for this partition function for any pure SYM theory is one of our main technical results. Perhaps surprisingly, our final answer \eqref{eq:Dpf} coincides with the partition function on the hemisphere background $HS^4$ obtained in \cite{Gava:2016oep, Dedushenko:2018tgx}. For the case of $SU(2)$ SYM it takes the form
\be
Z_{\text{AdS}}(a,\Lambda) = (\Lambda L)^{4 a^2} \, Z_{\text{1-loop}}(a) \, Z_{\text{Nekrasov}}(a, \Lambda) \, ,
\ee
where $Z_{\text{1-loop}}$ is the one-loop determinant for a vector multiplet with Dirichlet boundary conditions and $ Z_{\text{Nekrasov}}$ is the holomorphic part of the Nekrasov instanton partition function \cite{Nekrasov:2002qd} in the $\Omega$ background $\epsilon_1 = \epsilon_2 = 1/L$. While explicit, this answer is still rather involved: in order to perform the maximization we need to truncate the sum over instantons and use numerical techniques. 

As illustrated in figures \ref{fig:F} and \ref{fig:FF}, we find that
\begin{itemize}
\item{At weak coupling $\Lambda L \ll 1$, there is a single smooth local maximum of $\mathrm{Re}[F_\text{AdS}]$ at $\delta^* = 0$, corresponding to the $SU(2)$-preserving Dirichlet boundary condition;}
\item{As we increase $\Lambda L$, a new and larger local maximum of $\mathrm{Re}[F_\text{AdS}]$ emerges at $\delta^* \neq 0$. We take this to signal the transition to a new boundary condition with adjoint Higgsing;}
\item{As we further increase $\Lambda L$ to the largest values for which our numerical analysis is under control, we see yet another even larger maximum emerge. The difference in the values of $\mathrm{Re}[F_\text{AdS}]$ at the maxima becomes smaller as $\Lambda L$ grows. We conjecture that this pattern generalizes to a sequence of more and more maxima emerging and accumulating in the flat space limit $\Lambda L\to \infty$, from which the flat direction emerges.}
\end{itemize}
These results provide strong evidence that the expectations outlined above are met, though this happens through a rather intricate dynamical process that would have been hard to predict beforehand. Note that the transition to the Higgsed phase here does not happen through the continuous recombination process mentioned earlier, but rather it is a discontinuous ``first order'' type of transition. We should also clarify that, while we see the emergence of the new larger maxima, our analysis is not detailed enough to reveal any inconsistency in the $SU(2)$ preserving boundary condition at strong coupling.

As a byproduct, which lies somewhat outside of the main scope of the paper, we find a simple relation between the AdS partition function $F_{\text{AdS}}$, and a curved-space version of the prepotential $\mathcal{F}_{\text{AdS}}$, namely
\begin{equation}\label{eq:FisFintro}
F_{\text{AdS}} = -4\pi i \mathcal{F}_{\text{AdS}}~.
\end{equation}
Besides the fact that they are both holomorphic functions of $a$, this identification is motivated by: (i) the flat-space limit; (ii) the observation that the gauge-invariant VEV $u=-\langle \mathrm{Tr}[\phi^2]\rangle$ can be obtained from $F_{\text{AdS}}$ through a differential operator that, upon the identification \eqref{eq:FisFintro}, precisely reproduces the Matone relation\cite{Matone:1995rx}
\be \label{eq: matone}
u = \pi i \left( \cF(a) - \frac{1}{2} a \partial_a \cF(a)\right) \, ;
\ee
(iii) the observation that the condition $\mathrm{Im}\left[\mathcal{F}_{\text{AdS}}''\right]=\mathrm{Im}\left[\tau_{\text{IR}}\right] > 0$ matches the maximization condition $\partial_{\delta}^2\mathrm{Re}\left[F_{\text{AdS}}\right]<0$.

The plan for the rest of the paper is as follows: in \textbf{section 2} we discuss how to couple an $\cN=2$ vector multiplet to a background AdS$_4$ geometry with a manifest 3d $\mathcal{N}=2$ massive subalgebra, and we put forward our proposal of $F_{\text{AdS}}$ maximization to determine the boundary conditions preserving the full 3d $\mathcal{N}=2$ superconformal symmetry. In \textbf{section 3} we perform the localization calculation of the $\cN=2$ theory on an AdS background. In \textbf{section 4} we
leverage these results to discuss the boundary phase transitions for the $\cN=2$ $SU(2)$ SYM theory in AdS. In \textbf{section 5} we give a derivation of the AdS Matone relation and discuss the evidence for the identification \eqref{eq:FisFintro}. We conclude with comments about open directions for future research. 
Various details concerning the structure of SUSY multiplets, the localization computation and the numerical study of $F_{\text{AdS}}$-maximization are collected in appendices.

\section{4d $\mathcal{N}=2$ Super Yang-Mills in AdS}\label{s4dAdSsusy}
In this section we setup the problem by discussing the Lagrangian of $\mathcal{N}=2$ SYM in AdS$_4$, including boundary terms and 1/2-BPS boundary conditions that are needed to preserve a massive 3d $\mathcal{N}=2$ supersymmetry. Finally, we propose a criterion that selects boundary conditions preserving the full symmetry, isomorphic to a 3d $\mathcal{N}=2$ superconformal algebra.

\subsection{Lagrangian and boundary terms}
Let us start by reviewing how to put a 4d $\cN=2$ theory in rigid curved Euclidean spacetime, following mostly the conventions in \cite{Hama:2012bg}. We couple the theory to background supergravity, whose bosonic content is displayed in table \ref{Tab: weyl multiplet}.
\begin{table}[ht]
	\centering
	\begin{tabular}{cccc}
		\hline
		{Field}    & {Description}    & {$\text{U}(1)_{R}$} & {$\text{SU}(2)_{R}$ } \\ \hline
		$g_{\mu\nu}$       & metric                  & $0$                     & $\mathbf{1}$                  \\ 
		$\mathcal A_\mu$   & $\mathfrak{u}(1)_R$ gauge field   & $0$                     & $\mathbf{1}$                  \\ 
		$\mathcal V_\mu$   & $\mathfrak{su}(2)_R$ gauge field  & $0$                     & $\mathbf{3}$                       \\ 
		$T_{\mu\nu}$       & anti-self-dual tensor   & $+2$                    & $\mathbf{1}$                  \\ 
		$\bar{T}_{\mu\nu}$ & self-dual tensor        & $-2$                    & $\mathbf{1}$                  \\ 
		$\tilde{M}$        & scalar                  & $0$                     & $\mathbf{1}$                  \\  \hline
	\end{tabular}
	\caption{Bosonic field content of $\mathcal N=2$ supergravity multiplet.}\label{Tab: weyl multiplet}
\end{table}
We then fix the geometry, and fix the additional bosonic content of the supergravity multiplet in such a way that the Killing spinor equation admits solutions \cite{Festuccia:2011ws}. In order to obtain eight supercharges in Euclidean AdS$_4$, the bosonic content of the gravity multiplet is fixed to 
\begin{equation}\label{Eq: background sugra values}
	\tilde{M} = 0\:,\quad \mathcal A_\mu = 0 \:,\quad \mathcal V_\mu = 0\:,\quad  T_{\mu\nu} =0 \:,\quad  \bar T_{\mu\nu} = 0\:,\quad g_{\mu\nu} = g_{\mu\nu}^{\text{AdS}_4} \quad 
	\text{with}\quad\mathscr{R} = -12~,
\end{equation}
where the AdS radius is set to 1. Denoting the supersymmetry variation in background supergravity as
\begin{align}\label{eQ}
    \mathcal{Q}=\epsilon^A Q_A+\bar{\epsilon}_A\bar{Q}^A\,,
\end{align}
where $\epsilon$ and $\bar{\epsilon}$ are the spinorial parameter, and $Q_A$ and $\bar{Q}_A$ the operators acting on the fields, the Killing spinor equation is
\begin{align}\label{e4dN2AdSHSKs}
D_\mu \epsilon_{A\alpha}=-\frac i2 \tau_{3,A}^{\ \ B}(\sigma_\mu \bar{\epsilon}_B)_\alpha \,, \ \ D_\mu \bar{\epsilon}_A^{\dot{\alpha}}=\frac i2 \tau_{3,A}^{\ \ B}(\bar{\sigma}_\mu \epsilon_B)^{\dot{\alpha}} \,.
\end{align}
There are precisely eight linearly independent solutions, giving eight global supercharges. Our conventions for the spinors, the $\mathcal{R}$-symmetry indices, and the spacetime indices are detailed in the appendix \ref{stensorconv}. 
 
The field content of $\cN=2$ SYM theory is a 4d $\mathcal{N}=2$ vector multiplet, which comprises a complex scalar $\phi$, a Dirac fermion $\lambda_A$ in the doublet of $SU(2)_{\mathcal{R}}$, a gauge field $A_\mu$ and an auxiliary field $D_{AB}$ in the triplet of $SU(2)_{\mathcal{R}}$. All of these fields are in the adjoint representation of the gauge group $G_{\text{YM}}$. Specifying the general action for SYM coupled to background supergravity \cite{lauria2020supergravity} to the AdS$_4$ background, one gets the following bulk action 
\begin{align}\label{eactionAdS4}\allowdisplaybreaks
\begin{split}
S&=\frac{\text{Im}(\tau)}{4\pi}\int_{\text{AdS}_4}\,\mathcal L_{\text{YM}} + i\frac{\text{Re}(\tau)}{8\pi}  \int_{\text{AdS}_4} \,\mathcal L_{\theta}~,\\
\mathcal L_{\text{YM}} &=~ \text{Tr}\Big[\frac{1}{2}F^{\mu\nu}F_{\mu\nu}-4D_\mu\bar{\phi}D^\mu\phi+8\bar{\phi}\phi+i\bar{\lambda}^A\bar{\slashed{D}}\lambda_A-i\lambda^A\slashed{D}\bar{\lambda}_A\\
&\qquad\qquad -\frac{1}{2}D^{AB}D_{AB} +4[\phi,\bar{\phi}]^2-2\lambda^A[\bar{\phi},\lambda_A]+2\bar{\lambda}^A[\phi,\bar{\lambda}_A]\Big]\,,\\
    \mathcal L_{\theta} &= \text{Tr}\Big[\frac12 \varepsilon^{\mu\nu\rho\sigma} F_{\mu\nu} F_{\rho\sigma} \Big]~.
\end{split}
\end{align}
The background supergravity transformations of the vector multiplet, which can be specified to our background with \eqref{Eq: background sugra values}-\eqref{e4dN2AdSHSKs}, are given by \eqref{e4dvector}. The supersymmetry variation of the bulk Lagrangian is a total derivative. On a fixed geometry without a boundary the procedure above immediately gives a theory that preserves eight supercharges. However in AdS, in order to preserve at least part of the supersymmetry, we need to add boundary terms and consistent boundary conditions. To discuss the boundary term we introduce an IR cut-off $\eta_0$ in the radial coordinate, where the coordinate $\eta$ is defined by taking the following $S^3$ slicing of AdS$_4$
\begin{equation}\label{eq:coordinates}
ds^2=d\eta^2+\sinh(\eta)^2d\Omega_{S^3}^2\,.
\end{equation}
The cutoff breaks part of the isometries and consequently also part of supersymmetries (see e.g. \cite{Freedman:2012zz,Assel:2016pgi} for related discussions of supersymmetry in AdS). This means that only a subgroup of the full $OSp(2|4)$ AdS$_4$ supersymmetry algebra, acting as a 3d $\cN=2$ SUSY algebra on the boundary, is manifest in the formalism. For generic choices of the parameters that enter the boundary condition, only this subgroup is preserved, and thus the AdS isometries are broken. We discuss the additional constraints that are needed to preserve the full AdS$_4$ $\mathcal{N}=2$ algebra in section \ref{sFextremization}. 

More specifically, compatibly with our choice of cutoff in the coordinate $\eta$ in \eqref{eq:coordinates}, the largest supersymmetry algebra that can be made manifest from the Lagrangian supplemented by boundary terms is the massive $\mathcal{N}=2$ algebra on $S^3$. It is specified by the following constraint on the AdS$_4$ $\mathcal{N}=2$ Killing spinors 
\begin{align}\label{eq:e4dto3drestriction}
	\begin{split}
		&\epsilon_{A\alpha}=\sinh\left(\frac{\eta}{2}\right)\begin{pmatrix}
			\zeta_\alpha\\
			\tilde{\zeta}_\alpha
		\end{pmatrix}_A\,,\\
		&\bar{\epsilon}_{A\dot{\alpha}}=-i\cosh\left(\frac{\eta}{2}\right)\tau_{3,A}^{\ \ B}\begin{pmatrix}
			(\zeta\sigma_{\perp})_{\dot{\alpha}}\\
			(\tilde{\zeta}\sigma_{\perp})_{\dot{\alpha}}
		\end{pmatrix}_B\,,
	\end{split}
\end{align}
where $\zeta$ and $\tilde{\zeta}$ are Killing spinors on an $S^3$ of unit radius, see the appendix \ref{app: kspinor} for more details.

The boundary terms needed on the cutoff surface $\eta=\eta_0$ in order to preserve this subalgebra are
\begin{align}\label{ebractionAdS4}
\begin{split}
S_{\partial} &=  \frac{\text{Im}(\tau)}{4\pi} \sinh^3(\eta_0)\int_{S^3}\, \mathcal{L}_{\text{YM}}^{\partial} + i\frac{\text{Re}(\tau)}{8\pi} \sinh^3(\eta_0) \int_{S^3}\,\mathcal{L}_{\theta}^{\partial}~,\\
\mathcal{L}_{\text{YM}}^\partial=\frac{2}{\sinh(\eta_0)}&\left[\coth(\eta_0)\left(\frac{3+\cosh(2\eta_0)}{\sinh(\eta_0)}(\phi_1^2-\phi_2^2)-8i\coth(\eta_0)\phi_1\phi_2\right)\right.\\
&-\left(\cosh(\eta_0)\phi_2+i\phi_1\right)\left(4i\frac{D_\perp\phi_1}{\sinh(\eta_0)}+4\coth(\eta_0)D_\perp\phi_2+2iD_{12}\right)\\
&\left.+\cosh^2\left(\frac{\eta_0}{2}\right)\lambda_1\lambda_2+\sinh^2\left(\frac{\eta_0}{2}\right)\bar{\lambda}_1\bar{\lambda}_2\right]\,,\\
\mathcal{L}_{\theta}^\partial=\frac{2}{\sinh(\eta_0)}&\left[2\coth(\eta_0)\left(2\coth(\eta_0)(\phi_1^2-\phi_2^2)-\frac{3+\cosh(2\eta_0)}{\sinh(\eta_0)}i\phi_1\phi_2\right) \right.\\
&-\left(\phi_2+i\cosh(\eta_0)\phi_1\right)\left(4i\frac{D_\perp\phi_1}{\sinh(\eta_0)}+4\coth(\eta_0)D_\perp\phi_2+2iD_{12}\right)\\
&\left.+\cosh^2\left(\frac{\eta_0}{2}\right)\lambda_1\lambda_2-\sinh^2\left(\frac{\eta_0}{2}\right)\bar{\lambda}_1\bar{\lambda}_2-\frac{i}{2}\sinh(\eta_0)(\lambda_1\sigma_\perp\bar{\lambda}_2-\lambda_2\sigma_\perp\bar{\lambda}_1)\right]\,.
\end{split}
\end{align}
Here
\begin{align}\label{einsanedefinition}
    \phi_1=\frac{\phi+\bar{\phi}}{2i}\,,\ \ \ \phi_2=\frac{\phi-\bar{\phi}}{2}
\end{align}
and $\perp$ denotes the curved space index in the radial direction.

\subsection{Boundary Conditions} 
We now discuss boundary conditions at the cutoff surface $\eta=\eta_0$ that preserve the massive 3d $\mathcal{N}=2$ supersymmetry algebra. We will restrict ourselves to the case $\theta=0$ and describe the most basic Dirichlet and Neumann boundary conditions.\footnote{When the theta angle is nonzero, the Neumann boundary condition gets modified to a mixed boundary condition that fixes a linear combination of $\mathcal{J}_{\text{D}}^{(0)}$ and $\mathcal{J}_{\text{N}}^{(0)}$ in eq. \eqref{eJD&Ncurrentasymp}.} 
In the limit of large $\eta_0$, the possible modes for the bulk fields that are compatible with the linearized equation of motion, with the gauge-fixing $A_\perp = 0$, are
\begin{align}\label{easymptoticspergenera}
\begin{split}
\phi_{1,2}(\eta_0)&\underset{\eta_0\to\infty}{\sim} e^{-\eta_0}
\left(\phi_{1,2}^{(0)}+\hdots\right)+e^{-2\eta_0}\left(\tilde{\phi}_{1,2}^{(0)}+\hdots\right)\,,\\
\lambda(\eta_0)& \underset{\eta_0\to\infty}{\sim}e^{-\tfrac{3}{2}\eta_0}\left(\lambda_+^{(0)}+\hdots\right)+e^{-\tfrac{3}{2}\eta_0}\left(\lambda_-^{(0)}+\hdots\right)\,,\\
\bar{\lambda}(\eta_0)& \underset{\eta_0\to\infty}{\sim}e^{-\tfrac{3}{2}\eta_0}\left(\bar{\lambda}_+^{(0)}+\hdots\right)+e^{-\tfrac{3}{2}\eta_0}\left(\bar{\lambda}_-^{(0)}+\hdots\right)\,,\\
A_i(\eta_0)& \underset{\eta_0\to\infty}{\sim}\left(A_{i}^{(0)}+\hdots\right)+e^{-\eta_0}\left(\tilde{A}_{i}^{(0)}+\hdots\right)\,,
\end{split}
\end{align}
where the dependence on the $S^3$ coordinates is left implicit, and the boundary modes of the spinor satisfy $  \lambda_\pm^{(0)}=\pm i\sigma_\perp 
\bar{\lambda}_\pm^{(0)}$. A boundary condition is a choice of which of these coefficients are fixed at the boundary, and which are free to fluctuate, and we impose that this choice is compatible with supersymmetry. To this end, it is convenient to note that the 4d $\mathcal{N}=2$ multiplets of bulk operators can be decomposed, with $\eta$-dependent coefficients, in pairs of multiplets of the 3d $\mathcal{N}=2$ associated to the spinors $\zeta$ and $\tilde{\zeta}$ in \eqref{eq:e4dto3drestriction}. The 4d $\mathcal{N}=2$ vector multiplet of the elementary fields of SYM theory decomposes in the following two $\eta$-dependent 3d $\mathcal{N}=2$ multiplets 
\begin{align}\allowdisplaybreaks
\mathcal{J}_{\text{D}}(\eta):&
\begin{cases}
&J
=2\phi_1\,,\\
&j=-{\displaystyle\frac{i}{\sqrt{2}}}\big(\sinh\big(\tfrac{\eta}{2}\big)\lambda_2+i\cosh\big(\tfrac{\eta}{2})\sigma_\perp\bar{\lambda}_2\big)\,,\\
&\tilde{j}={\displaystyle \frac{i}{\sqrt{2}}}\big(\sinh\big(\tfrac{\eta}{2}\big)\lambda_1-i\cosh\big(\tfrac{\eta}{2})\sigma_\perp\bar{\lambda}_1\big)\,,\label{eadsjd}\\
&j_k=-{\displaystyle\frac{i}{\sinh^2(\eta)}}\left(\cosh(\eta)\tfrac 12\varepsilon^{ij}_{\ \ k}F_{ij}+\sinh(\eta)F_{k\perp}\right)\,,\\
&K=-2\sinh(\eta)D_\perp\phi_2-2\cosh(\eta)(\phi_2+\tfrac{i}{2}D_{12})+2i\phi_1\,.
\end{cases} \\
\mathcal{J}_{\text{N}}(\eta):&
\begin{cases}
&J=2\phi_2\,,\\
&j={\displaystyle\frac{1}{\sqrt{2}}}\big(\sinh\big(\tfrac{\eta}{2}\big)\lambda_2-i\cosh\big(\tfrac{\eta}{2})\sigma_\perp\bar{\lambda}_2\big)\,,\\
&\tilde{j}=-{\displaystyle\frac{1}{\sqrt{2}}}\big(\sinh\big(\tfrac{\eta}{2}\big)\lambda_1+i\cosh\big(\tfrac{\eta}{2})\sigma_\perp\bar{\lambda}_1\big)\,,\label{eadsjn}\\
&j_k=-{\displaystyle\frac{1}{\sinh^2(\eta)}}\left(\tfrac 12\varepsilon^{ij}_{\ \ k}F_{ij}+\cosh(\eta)\sinh(\eta)F_{k\perp}\right)\,,\\
&K=2\sinh(\eta)D_\perp\phi_1+2\cosh(\eta)\phi_1-D_{12}+2i\phi_2\,.
\end{cases}
\end{align}
We recall the supersymmetry transformations of the 4d and 3d multiplets in the appendix \ref{asusytransf}.
The multiplets $\mathcal{J}_{\text{D}}$ and $\mathcal{J}_{\text{N}}$ contain a current $j_k$ in the adjoint representation of the gauge group that is covariantly conserved. Therefore, compared to a more standard linear multiplet of a conserved current, their supersymmetry variation contain additional terms that involve the following 3d $\mathcal{N}=2$ vector multiplet, which is again built from the 4d vector multiplet using $\eta$-dependent coefficients
\begin{equation}\label{eads}   
\mathcal{V}(\eta):
\begin{cases}
&\sigma=2\cosh(\eta)\phi_1-2i\phi_2\,,\\
&\lambda=-\sinh(\eta)\left(\cosh\big(\frac \eta 2\big)\lambda_1-i\sinh\big(\frac \eta 2\big)\sigma_\perp\bar{\lambda}_1\right)\,,\\
&\tilde{\lambda}=-\sinh(\eta)\left(\cosh\big(\frac \eta 2\big)\lambda_2+i\sinh\big(\frac \eta 2\big)\sigma_\perp\bar{\lambda}_2\right)\,,\\
&A_i=A_i\,,\\
&D=-2\cosh(\eta)\sinh(\eta)D_\perp\phi_2-2i\sinh(\eta)D_\perp \phi_1-i\sinh^2(\eta)D_{12}\\
&\ \ \ \ \ \ +2\sinh^2(\eta)\phi_2-i\cosh(\eta)\phi_1-\phi_2\,.
\end{cases}
\end{equation}
Having built these multiplets, the boundary conditions 
\begin{align}
\begin{split}
\text{Dirichlet:}&~~\lim_{\eta_0\to\infty} e^{\eta_0}\mathcal{J}_{\text{D}}(\eta_0) = 0~,\\
\text{Neumann:}&~~\lim_{\eta_0\to\infty} e^{\eta_0}\mathcal{J}_{\text{N}}(\eta_0) = 0~,
\end{split}
\end{align}
manifestly preserve the massive 3d $\mathcal{N}=2$ supersymmetry on $S^3$. Written in terms of the boundary modes in equation \eqref{easymptoticspergenera}, they set to zero either one of the two following linear multiplets on $S^3$ 
\begin{align}
\begin{split}
&\mathcal{J}_{\text{D}}^{(0)}=(2\phi_1^{(0)},-\sqrt{2} i \lambda_{2\,+}^{(0)},\sqrt{2} i \lambda_{1\,-}^{(0)},-i \varepsilon_i^{ jk}\partial_jA_k^{(0)},2(\tilde{\phi}_2^{(0)}+ i \phi_1^{(0)}))\,,\label{eJD&Ncurrentasymp}\\
&\mathcal{J}_{\text{N}}^{(0)}=(2\phi_2^{(0)},\sqrt{2}\lambda_{2\,-}^{(0)},-\sqrt{2}\lambda_{1\,+}^{(0)},-\tilde{A}_i^{(0)},-2( \tilde{\phi}_1^{(0)}-i\phi_2^{(0)}))\,.
\end{split}
\end{align}
Here we are using the notation $\mathcal{J}=(J,j,\tilde{j},j_i,K)$ for the components of the linear multiplet. We used that the auxiliary field $D_{12}$ is set to zero by the equations of motion, so there is no associated boundary mode. Note that the current $j_i$ that we are setting to zero is $-i \varepsilon_i^{ jk}\partial_jA_k^{(0)}$ in the case of $\mathcal{J}_{\text{D}}^{(0)}$, and $\tilde{A}_i^{(0)} = -\underset{\eta\to\infty}{\lim}e^{\eta}F_{\perp i}$ in the case of $\mathcal{J}_{\text{N}}^{(0)}$, so these boundary conditions are the supersymmetrizations of the ordinary Dirichlet and Neumann boundary conditions for a gauge field, and are thus guaranteed to define a good variational principle. Since with Dirichlet the connection trivializes at the boundary, the multiplet $\mathcal{J}_{\text{N}}^{(0)}$ of fluctuating modes is an ordinary linear multiplet with a conserved current $j_i$, and we have a global symmetry $G_{\text{YM}}$ at the boundary. On the other hand, with Neumann, $\mathcal{J}_{\text{D}}^{(0)}$ is still the multiplet of a covariantly-conserved current, coupled to the boundary limit of the vector multiplet $\mathcal{V}$, and the symmetry $G_{\text{YM}}$ is still gauged at the boundary.

We can allow a more general, ``modified Dirichlet'' boundary condition that preserves the same supersymmetry, namely
\begin{align}\label{edirichletbc}
\mathcal{J}_{\text{D}}^{(0)}=(-a,0,0,0,0)\,,
\end{align}
where $a$ is an adjoint-valued constant, that corresponds to turning on a source for the top component of the fluctuating linear multiplet $\mathcal{J}_{\text{N}}^{(0)}$. If we do so, however, it is possible to see from \eqref{econstraintslinear} that the conserved current at the boundary breaks to the Cartan spanned by $a$.

\subsection{$F_{\text{AdS}}$-maximization}\label{sFextremization}

As explained in the introduction, our goal is studying the transition at strong coupling between the non-abelian Coulomb phase, i.e. $SU(2)$ preserving Dirichlet boundary condition, and the phase Higgsed down to $U(1)$. In order to properly set up the problem, and find that some boundary conditions stop being available at strong coupling, we need to be careful in specifying what set of boundary conditions we consider allowed. To clarify this point, consider the example of non-supersymmetric Yang-Mills \cite{Aharony:2012jf, Ciccone:2024guw}: in that context, the leading scenario for the transition to the confinement phase is that the Dirichlet boundary condition disappears through a mechanism of merger and annihilation. This does not mean that no boundary condition of Dirichlet-type is allowed when AdS is large, it just means that such a boundary condition is incompatible with the requirements of unitarity and invariance under AdS isometries, i.e. conformal symmetry at the boundary. Also in the present setup we expect that it is important to restrict to the set of boundary condition that realize the full (super)group of AdS (super-)isometries. In the previous section we saw that only a 3d $\mathcal{N}=2$ massive supersymmetry algebra on $S^3$ is manifest in the action that we use. We then need to find additional conditions that ensure that the boundary condition actually preserves the full symmetry, i.e. 4d $\mathcal{N}=2$ supersymmetry in AdS$_4$, acting on the boundary as the 3d $\mathcal{N}=2$ superconformal algebra. In this section we conjecture what these conditions are, drawing inspiration from the analogous problem that one encounters for  supersymmetric quantum field theory on $S^3$, namely ensuring that the curvature couplings are fixed in such a way to preserve superconformal symmetry.

Let us recall how this works for 3d theories \cite{Jafferis:2010un, Closset:2012vg}. The 3d $\mathcal{N}=2$ supersymmetry algebra in flat space has an outer automorphism $U(1)_{\mathcal{R}}$ which can be a global symmetry of the theory. If the theory admits additional global symmetries $U(1)_a$, the $U(1)_{\mathcal{R}}$ can be redefined adding to its generator arbitrary linear combinations of those of the $U(1)_a$'s, namely
\begin{align}\label{edeltaR}
\mathcal{R} = \mathcal{R}_0 +\sum_a \delta^a \, q_a \,,   
\end{align}
where $\mathcal{R}_0$ is an arbitrary $\mathcal{R}$-charge assignment consistent with the Lagrangian, and $q_a$ is the charge of $U(1)_a$. Theories that admit a $U(1)_{\mathcal{R}}$ symmetry can be put on $S^3$ by coupling to the new minimal background supergravity multiplet \cite{Closset:2012ru}. The resulting curvature couplings, the supersymmetry algebra on $S^3$, and the value of the $S^3$ partition function, all depend on which assignment of $\mathcal{R}$-charge is chosen among the possibilities in \eqref{edeltaR}. If the starting theory in flat space is superconformal, there is another way to put it on $S^3$ that preserves the full superconformal symmetry, which is to simply perform a Weyl rescaling. The curvature couplings obtained in this way match those in the previous construction only for a special choice of the $\mathcal{R}$-charge assignment: namely if one picks the particular $\mathcal{R}$-symmetry that belongs to the superconformal algebra, i.e. the so-called superconformal $\mathcal{R}$-symmetry. The latter is not subject to any (continuous) ambiguity because its charge can be obtained from anti-commutators of the odd superconformal generators $\{\mathcal{Q},\mathcal{S}\}= \mathcal{R}_\text{SCFT}+\dots$~. It was proposed in \cite{Jafferis:2010un}, and later proved in \cite{Closset:2012vg}, that the value of the mixing parameters $\delta^a$ that give the superconformal $\mathcal{R}$-charge is the one that maximizes the quantity $\mathrm{Re}[F_{S^3}] = - \log |Z_{S^3}|$, where $F_{S^3}= -\log Z_{S^3}$ is the partition function obtained by putting the theory on the sphere using the generic $\mathcal{R}$-symmetry.

Going back to AdS$_4$, the bulk $\mathcal{N}=2$ algebra is isomorphic to the 3d $\mathcal{N}=2$ superconformal algebra, in particular it contains an unambiguous bulk $U(1)_\mathcal{R}$ symmetry with fixed charge assignment $\mathcal{R}_{\text{AdS}}$. However, the boundary condition at the cutoff surface $\eta=\eta_0$ is only 1/2-BPS, and the preserved $U(1)_\mathcal{R}$ is allowed to mix with other $U(1)$ charges like in \eqref{edeltaR}. In AdS the parameters $\delta^a$ should be thought of as parametrizing different possible 1/2-BPS boundary conditions at the cutoff surface $\eta=\eta_0$. We conjecture that also in this setup the value of $\delta^a$ that preserves the full bulk supersymmetry is the one that maximizes the real part of a supersymmetric partition function, in this case the AdS one
\begin{align}\label{eFAdS}
   F_{\text{AdS}}=-\log\,Z_{\text{AdS}}\,,
\end{align}
where, similarly to the case of $S^3$, $Z_{\text{AdS}}$ is computed with the 1/2-BPS boundary condition and it is a function of the parameters $\delta^a$. The way that the parameters $\delta^a$ enter in $F_{\text{AdS}}$ can be understood from the general observation of \cite{Jafferis:2010un}: for a theory with 3d $\mathcal{N}=2$ supersymmetry, the partition function depends holomorphically on the combination $m^a-i\delta^a$, where $m^a$ is a real mass deformation associated to the global symmetry $U(1)_a$. In the context of AdS, this means that the mixing can happen with global symmetries that act locally on the boundary, i.e. having an associated conserved current at the boundary, in terms of which one can define a real mass deformation. One important class of examples with boundary conserved currents, which will be our main focus, is that of gauge theories with Dirichlet boundary conditions. In that case, the real mass deformation at the boundary corresponds to a Cartan-valued parameter entering the modified Dirichlet boundary condition \eqref{edirichletbc}, i.e. defining $a= a^a H_a$, with $H_a$ the Cartan generators, we have $a^a= m^a$. Therefore, the mixing with the $R$-symmetry is achieved by complexifying the parameter in the boundary condition and considering $a^a = m^a - i \delta^a$.

Besides the mere analogy with the $S^3$ case, the following arguments support the conjecture of $F_{\text{AdS}}$-maximization:
\begin{itemize}
\item {\bf SCFT on AdS$_4$:} Consider the special case in which the bulk theory is a SCFT. A SCFT in AdS$_4$ background with a boundary condition that preserves the full supersymmetry can be mapped, through a Weyl rescaling, to the hemisphere $HS^4$, with a superconformal boundary condition on the equatorial $S^3$. Therefore, when the bulk theory is superconformal, we have a one-to-one correspondence between AdS$_4$ invariant boundary conditions and superconformal boundary conditions on $HS^4$. When the SCFT is placed on the hemisphere $HS^4$ with a boundary condition on the equatorial $S^3$, the same ambiguity as \eqref{edeltaR} arises for the $U(1)_{\mathcal{R}}$ symmetry preserved by the boundary condition. In \cite{Gaiotto:2014gha} it was argued that in order to find superconformal boundary conditions one needs to determine the values of $\delta^a$'s that maximize the real part of the partition function $F_{HS^4} = - \log Z_{HS^4}$.\footnote{Reference \cite{Gaiotto:2014gha} further normalizes $|Z_{HS^4}|$ dividing it by a factor of $|Z_{S^4}|^{\frac{1}{2}}$. This ensures the cancellation of contributions due to bulk Weyl anomalies or ambiguous curvature counterterms. The resulting combination $F_\partial = - \log |Z_{HS^4}|/(|Z_{S^4}|^{\frac{1}{2}})$ is a universal quantity that decreases along boundary RG flows, as was proved using entropic arguments in \cite{Casini:2018nym}. The normalization is however not needed to implement the maximization: bulk Weyl anomalies or local counterterms cannot depend on the parameters $\delta^a$ of the boundary conditions. Similar remarks on the dependence on Weyl anomalies and ambiguous counterterms apply also to AdS.} 
Then, our proposal coincides with the one \cite{Gaiotto:2014gha} if $F_{\text{AdS}}=F_{HS^4}$, so that maximizing this quantity to find AdS$_4$ invariant boundary condition is the same as the procedure of \cite{Gaiotto:2014gha} to find superconformal boundary conditions. The relation $F_{\text{AdS}}=F_{HS^4}$ trivially holds at the extremal value of $\delta^a$, because in that case the two partition functions are related by the Weyl rescaling mentioned before. However we need this relation to hold also for general $\delta^a$'s in order to support our conjecture. As an evidence for this, in the next section we perform a localization calculation of $Z_{\text{AdS}}$ for pure SYM theory, allowing for generic mixing of the $U(1)_\mathcal{R}$, and we find a match with the result for $Z_{HS^4}$ obtained in \cite{Gava:2016oep, Dedushenko:2018tgx}. While we have not performed the full localization calculation in AdS for generic gauge theories with matter, we have checked that the one-loop determinants are the same on AdS and $HS^4$ even when hypermultiplet matter is included, in particular for superconformal gauge theories. From this perspective, our proposal of $F_{\text{AdS}}$ maximization can be seen as an extension of the proposal of \cite{Gaiotto:2014gha} from the special case of SCFTs to massive QFTs in AdS$_4$.
\item {\bf Holography:} If we consider a 3d SCFT that has an AdS$_4$ supergravity dual, turning on generic $U(1)_\mathcal{R}$ mixing in the SCFT maps to boundary conditions that are not invariant under the full AdS$_4$ isometry on the supergravity side. The equality $Z^{\text{SCFT}}_{S^3} = Z_{\text{AdS}}$ between the deformed partition functions is a basic entry of the holographic dictionary. Therefore in this context the boundary conditions that preserve the full 3d $\mathcal{N}=2$ algebra are manifestly obtained by maximizing $\mathrm{Re}[F_{\text{AdS}}]$. From this perspective, our proposal arises by taking the rigid limit $M_{\text{Pl}}\to \infty$ to recover a QFT in AdS. See references \cite{Freedman:2013oja,Zan:2021ftf, Binder:2021euo} for holographic calculation in AdS$_4$ duals of the partition function of 3d SCFTs. As a further example of the predictive power of this decoupling limit for QFT in AdS, note that \cite{Zan:2021ftf} found a direct relation between $F_{S^3}$, seen as a function of the mixing parameters, and the prepotential of the supergravity theory, as a function of the scalars in vector multiplets. As we discuss in section \ref{sec:F=F} below, the quantity $F_{\text{AdS}}$, as a function of the boundary values of the scalar fields, can in fact be seen as a fully quantum-corrected prepotential for the theory in AdS background, consistently with the decoupling limit.
\end{itemize}

\section{AdS$_4$ localization}\label{sAdSlocalization}
In this section we perform the localization computation of $F_{\text{AdS}}= -\log Z_{\text{AdS}}$ for SYM. This will be the crucial quantity to discuss the transition to the Higgs phase for $G_\text{SYM} = SU(2)$ in section \ref{snewmaxima}. Readers that are only interested in this application can simply read the result for $Z_{\text{AdS}}$ in equation \eqref{eq:Dpf} and skip to the next section. Localization in AdS spacetime has been studied in \cite{Dabholkar:2014wpa,Assel:2016pgi,David:2016onq,David:2019ocd,David:2018pex,GonzalezLezcano:2023cuh, Iannotti:2023jji, BenettiGenolini:2023ndb, BenettiGenolini:2024lbj, BenettiGenolini:2024xeo}.

The theory we are considering has a marginally-relevant gauge coupling, with one-loop exact beta function,  and this is seen in the localization calculation from a renormalization needed to absorb UV divergences in the one loop determinants \cite{Pestun:2007rz} (we show the details in the appendix \ref{a1loopads}). As a result the coupling is traded with a scale $\mu e^{\frac{2\pi i\tau(\mu)}{b_1}} = \Lambda $ where $b_1$ is the coefficient of the one-loop beta function, for pure $SU(2)$ SYM $b_1=4$. In particular taking $\tau$ to be the coupling at the scale of the inverse AdS radius we have
\begin{equation}\label{ecouplingren}
e^{\pi i\tau} = (\Lambda L)^{\frac{b_1}{2}}~,
\end{equation}
where we restored $L$ for clarity, though we will keep $L=1$ in the calculation. The partition functions are written in terms of $\Lambda$ and the Lie algebra matrix $a$ in equation \eqref{bcloc}, which is fixed with Dirichlet boundary condition, and fluctuating with Neumann boundary condition. In both cases there are one loop contributions from the determinants, expressed in terms of Barnes $G$-functions through the combination
\begin{equation}
	H(x) = G(1+x) G(1-x)\,.
\end{equation}
and nonperturbative contributions that are encoded by the Nekrasov instanton partition function $Z_{\text{Nekrasov}}(a,\Lambda)$ \cite{Nekrasov:2002qd}. In both cases the final result involves a product over the positive roots $\Delta_+$ of the gauge group $G_{\text{YM}}$. The results are

\begin{itemize}
\item \textbf{Dirichlet boundary conditions}
\begin{equation}\label{eq:Dpf}
	Z^{\text{D}}_{\text{pure SYM}}(a,\Lambda) = \Lambda^{\frac{b_1}{2} \text{Tr}(a^2)}\prod\limits_{\alpha \in \Delta^+} H(i\alpha\cdot a) \frac{\alpha\cdot a}{\sinh (\pi \alpha\cdot a)}Z_{\text{Nekrasov}}(a,\Lambda)\,.
\end{equation}

\item \textbf{Neumann boundary conditions}
\begin{align}
\mathcal Z^{\text{N}}_{\text{pure SYM}} (\Lambda)=\int\!d a\ \Lambda^{\frac{b_1}{2} \text{Tr}(a^2)}\prod\limits_{\alpha \in \Delta^+}(\alpha\cdot a) \sinh(\pi \alpha \cdot a)H(i\alpha\cdot a) \cdot Z_{\text{Nekrasov}}(a,\Lambda) \, .
\end{align}
Here the integral over the Lie algebra has been rewritten as an integral over the Cartan sub-algebra including the contribution from the Vandermonde determinant.
\end{itemize}
We have picked the following Killing spinor to define the localization charge $\mathcal{Q}_{\text{loc}}$
\bea\label{eKillingloc}
&\epsilon_{A\alpha}=\frac{1}{\sqrt{2}} \sinh\bigg(\frac \eta2\bigg) \delta_{A\alpha}\,,\ \ \ \bar{\epsilon}_A^{\dot{\alpha}}=\frac{i}{\sqrt{2}}\cosh\bigg(\frac \eta2\bigg) \tau_{3, A}^{\ \ \alpha}\,.
\eea
We proceed by studying the locus together with the classical contribution. We then discuss the instanton corrections, leaving the 1-loop determinant computations to the appendix \ref{a1loopads}. Note that to perform the calculation of the one-loop determinants we use a fixed point formula \eqref{eindex}, commonly used in the literature for this type of calculations. This formula applies rigorously only on compact spaces, but it was explicitly shown, for instance in \cite{GonzalezLezcano:2023cuh, Assel:2016pgi}, that when this formula is applied to non-compact hyperbolic manifolds the final result is consistent with the one obtained by other methods. To make the presentation of the localization calculation less involved, in most of this section we skip the details of the gauge-fixing. We comment the role of the gauge-fixing in section \ref{sgaugefixing}. 

\subsection{The AdS locus and classical contributions}\label{slocus}
To perform localization we deform the action by a $\mathcal{Q}_{\text{loc}}$-exact term $S_{\text{SYM}} + t \int_{\text{AdS}}\mathcal{Q}_{\text{loc}}V$ and take the limit $t\to +\infty$. We take
\be
V = \text{Tr}\big[\lambda^A(\cQ_{\text{loc}} \lambda^A)^*\big]\,.
\ee
Here the ${}^*$ denotes complex conjugation, which is defined on the Euclidean fields as
\begin{align}
	(\phi)^*=-\bar{\phi}\,,\ \ (A_\mu)^*=A_\mu\,,\ \ (D_{AB})^*=-D^{AB}\,.
\end{align}
This choice ensures that the action \eqref{eactionAdS4} is a sum of positive terms, so that the path integral is convergent. 

The resulting localization action for the bosonic fields takes the form
\begin{align}\label{eQVinitial}
\begin{split}
\mathcal{Q}_{\text{loc}}V\big|_{\text{bosonic}}= \text{Tr}\bigg[&\epsilon^A\epsilon_A\bigg(F^-_{\mu\nu}+\frac{2}{\epsilon^A\epsilon_A}v_{[\mu}D_{\nu]^-}\phi_2+ i\frac{\epsilon^A\sigma_{\mu\nu}\epsilon^B\tau_{3,AB}}{\epsilon^C\epsilon_C}\phi_2\bigg)^2\\
&+\bar{\epsilon}_A\bar{\epsilon}^A\bigg(F^+_{\mu\nu}-\frac{2}{\bar{\epsilon}_A\bar{\epsilon}^A}v_{[\mu}D_{\nu]^+}\phi_2- i\frac{\bar{\epsilon}^A\bar{\sigma}_{\mu\nu}\bar{\epsilon}^B\tau_{3,AB}}{\bar{\epsilon}_C\bar{\epsilon}^C}\phi_2\bigg)^2\\
&+\frac{4}{\epsilon^A\epsilon_A+\bar{\epsilon}_A\bar{\epsilon}^A}\big(D_\mu[(\epsilon^A\epsilon_A+\bar{\epsilon}_A\bar{\epsilon}^A)\phi_1]\big)^2+\bigg(\frac{1}{\epsilon^A\epsilon_a}+\frac{1}{\bar{\epsilon}_A\bar{\epsilon}^A}\bigg)(v\cdot D\phi_2)^2\\
&+\frac{1}{2(\epsilon^A\epsilon_A+\bar{\epsilon}_A\bar{\epsilon}^A)}\big|2\phi_1(\epsilon^C\epsilon_C-\bar{\epsilon}_C\bar{\epsilon}^C)\tau_{3,AB}+(\epsilon^C\epsilon_C+\bar{\epsilon}_C\bar{\epsilon}^C)D_{AB}\big|^2\\
&-16(\epsilon^A\epsilon_A+\bar{\epsilon}_A\bar{\epsilon}^A)[\phi_1,\phi_2]^2\bigg]\,,
\end{split}
\end{align}
where the $\pm$ superscript refers to the self-dual/anti-self-dual components of the tensor, and $v^\mu =2\bar{\epsilon}^A\bar{\sigma}^\mu\epsilon_A$. This action is a sum of squares of real quantities, in particular it is positive semi-definite, and therefore the stationary point in the limit $t\to+\infty$ is given by $\mathcal{Q}_{\text{loc}}V = 0$, which defines the localization locus. 

In order to derive the locus, we make a simplifying assumption: we assume that the field configuration at the locus is invariant under the $SO(4)$ symmetry that acts on the $S^3$ at fixed radial coordinate $\eta$. This means that we allow only $\eta$-dependent profiles for each field, and we look for the locus among these configurations. In particular, while $SO(4)$ invariance still allows a nonzero $A_\eta$ which is only function of $\eta$, that can be always set to zero by a gauge transformation. As a result the gauge connection trivializes --with the important exception of singular instanton contributions localized at the origin, as we discuss below-- and the gauge-covariant derivatives can be treated as ordinary derivatives. It is convenient to rewrite the expression \eqref{eQVinitial} as
\begin{align}\label{QVAdS}
\begin{split}
\mathcal{Q}_{\text{loc}}V\big|_{\text{even}}=\text{Tr}\bigg[& \frac{4}{\cosh(\eta)}\big(D_\mu[\cosh(\eta)\phi_1]\big)^2 +\frac{1}{2\cosh(\eta)}\big|-2\phi_1\tau_{3,AB}+ \cosh(\eta) D_{AB}\big|^2\\
& +4\phi_2\big(-D^2-2\big)\phi_2 -16\cosh(\eta)[\phi_1,\phi_2]^2\\
&+\cosh^2\left(\frac{\eta}{2}\right)\bigg(F^-_{\mu\nu}+ i\frac{\epsilon^A\sigma_{\mu\nu}\epsilon^B\tau_{3,AB}}{\epsilon^C\epsilon_C}\phi_2\bigg)^2\\ & + \sinh^2\left( \frac{\eta}{2} \right)\bigg(F^+_{\mu\nu}- i\frac{\bar{\epsilon}^A\bar{\sigma}_{\mu\nu}\bar{\epsilon}^B\tau_{3,AB}}{\bar{\epsilon}_C\bar{\epsilon}^C}\phi_2\bigg)^2\\
&\,+D_\nu\bigg(D^\nu\big(2\cosh(\eta)\phi_2^2\big)-4\tilde{F}^{\mu\nu}v_\mu\phi_2\bigg)\bigg]\,,
\end{split}
\end{align} 
where $\tilde{F}_{\mu\nu}=\tfrac 12 \varepsilon_{\mu\nu\rho\sigma}F^{\rho\sigma}$. Unlike \eqref{eQVinitial}, this rewriting is not a sum of squares of real quantities, but we still obtain a sum of non-negative terms up to a total derivative. Indeed $-D^2-2$ has a positive spectrum, because the possible eigenvalues of the scalar Laplacian $D^2$ are smaller than the Breitenlohner-Freedman bound $m^2_{\text{BF}}L^2  = -\frac{9}{4}$ \cite{Breitenlohner:1982jf}. We can then proceed to find the locus by imposing that each term vanishes. Setting to zero the terms in the first line gives
\begin{align}
\begin{split}\label{ephi1locus}
\phi_1 = -\frac{a}{2\cosh(\eta)} \, , ~
D_{AB} = - \frac{1}{\cosh^2(\eta)} \, \tau_{3, AB} \, a \, .
\end{split}
\end{align}
with $a$ a constant Lie-algebra valued matrix. Setting to zero the terms in the second line gives
\begin{equation}
\phi_2 = \frac{c_1}{\sinh(\eta)^2}+ \frac{c_2 \cosh(\eta)}{\sinh(\eta)^2}~,~~[c_1,a]=[c_2,a]=0~.
\end{equation}
Setting to zero the term in the third line, since the connection is trivial, forces
\begin{equation}
c_1 = c_2 = 0~.
\end{equation}
This automatically satisfies also the constraint given by the term in the fourth line. In addition, since the coefficient of this term is $\sinh(\eta)^2$, we can also allow for a discontinuous field configuration that is non-vanishing only at the center $\eta=0$. The field $\phi_2$ solves the classical equation of motion everywhere, to ensure vanishing of the term in the second line, so it cannot have such a discontinuity. However, by analogy with the $S^4$ case \cite{Pestun:2007rz}, we can allow singular instantons for which $F^+_{\mu\nu}$ is non-vanishing at the center, while $F^-_{\mu\nu} = 0$. We discuss their contributions below in section \ref{ainstantons}. Note that also the total derivative term vanishes identically when we plug the bulk solution $\phi_2 = 0$.

The field configuration that we obtained in this way satisfies
\begin{align}
\begin{split}\label{bcloc}
\mathcal{J}_{\text{D}}^{(0)}& = \lim_{\eta_0\to\infty} e^{\eta_0}\mathcal{J}_{\text{D}}(\eta_0) = (-a,0,0,0,0)~,\\
\mathcal{J}_{\text{N}}^{(0)}& =\lim_{\eta_0\to\infty} e^{\eta_0}\mathcal{J}_{\text{N}}(\eta_0) = 0~,
\end{split}
\end{align}
so it is compatible with either the choice of (modified) Dirichlet or of Neumann boundary condition defined above. For (modified) Dirichlet the matrix $a$ is interpreted as a fixed parameter of the boundary condition, while for Neumann it belongs to the multiplet that is fluctuating at the boundary, therefore one still needs to perform a finite-dimensional integral over it. There is however an observation to be made about the condition on $\mathcal{J}_{\text{D}}^{(0)}$: while the component $K$ of $\mathcal{J}_{\text{D}}^{(0)}$ is vanishing both for $t=0$ (i.e. the original SYM action) and for $t\to\infty$, consistently with supersymmetry, we observe that its expression in terms of the modes of the component fields at the boundary is different in the two cases. As we showed above, at $t=0$ we have $K=2(\tilde{\phi}_2^{(0)}+ i \phi_1^{(0)})$, so to ensure $K=0$ we need $\tilde{\phi}_2^{(0)} = -i \phi_1^{(0)} = i \frac{a}{2}$. On the other hand, at the locus the field $\phi_2$ vanishes identically, but the non-trivial background \eqref{ephi1locus} for the auxiliary field $D_{12}$, when plugged in $e^{\eta_0}\mathcal{J}_{\text{D}}(\eta_0)$, ensures that there is again a cancellation of the $K$ component. The fact that at the localization locus there is a nonzero value for an auxiliary field of the vector multiplet, proportional to a scalar in the vector multiplet, is actually common in localization, see e.g. \cite{Pestun:2007rz, Gava:2016oep, Dedushenko:2018tgx, Assel:2016pgi}. Therefore both (modified) Dirichlet and Neumann boundary conditions can be expressed in the same way in terms of the boundary multiplets $\mathcal{J}_{\text{D}}^{(0)}$ and $\mathcal{J}_{\text{N}}^{(0)}$ at $t=0$ and $t\to\infty$. It is natural to expect that this is a general consequence of the fact that these conditions remain valid at arbitrary values of $t$. In order to check this one would need to study the variational principle for $\mathcal{L}_{\text{SYM}} + t \mathcal{Q}_{\text{loc}}V$.

The classical contribution to the localization formula is given by plugging the locus \eqref{ephi1locus} in the action \eqref{eactionAdS4}-\eqref{ebractionAdS4}. The result it
\begin{align}\label{eclassical}
    e^{-S}\bigg|_{\text{loc}}=e^{\pi i\tau\text{Tr}[a^2]}\,.
\end{align}
We then trade $e^{i\pi\tau}$ to $\Lambda^{\frac{b_1}{2}}$ as explained in \eqref{ecouplingren}.

\subsection{Instantons in AdS$_4$}\label{ainstantons}

As we said above, the minimization of the localization action allows discontinuous configurations in which $F^+_{\mu\nu}$ is non-vanishing only at the center $\eta=0$. These configurations satisfy $F^-_{\mu\nu}=0$ and are localized instantons, namely as in the case of $S^4$ \cite{Pestun:2007rz} we include in the path integral singular configurations of the gauge field. The fact that these configurations are localized allows to escape the no-go result for instanton contributions in the presence of Dirichlet boundary conditions \cite{Callan:1989em, Aharony:2012jf}.

The $\hat{\mathcal{Q}}_{\text{loc}}$-complex, for the part of the path integration close to the center of AdS$_4$, is equivalent to the one of \cite{Nekrasov:2002qd}. As for $S^4$ case we therefore obtain a Nekrasov partition function contribution. The Killing vector generated by $\hat{\mathcal{Q}}_{\text{loc}}^2$ fixes the parameters $\epsilon_1, \, \epsilon_2$ of the $\Omega$ background to be
\be
\epsilon_1 = \epsilon_2 = \frac{1}{L} \, .
\ee

The sum is just over the instanton contributions, and not anti-instantons. Indeed, as opposed to the $S^4$ case and similarly to the $HS^4$, we only have one point where the field configuration can be singular while $F^-_{\mu\nu}=0$. The difference between instantons and anti-instantons comes from the choice of orientation of the boundary. Note that the full dependence on $\tau$ comes from the classical and the nonperturbative contribution to the partition function, and both these contributions are holomorphic in $\tau$. A change of the boundary orientation would make it anti-holomorphic.

\section{Transition to adjoint-Higgsing in pure $SU(2)$ SYM}\label{snewmaxima}
We are now ready to apply the results of the previous sections to study the large radius / strong coupling limit of pure $SU(2)$ SYM in AdS$_4$, starting with the Dirichlet boundary condition with $SU(2)$ boundary global symmetry at weak coupling. We study this problem via the the maximization of $F_{\text{AdS}}$ as a function of the complexified parameter $a= -i \delta$, which ensures that the boundary condition satisfies AdS (super-)isometries. More precisely, we will maximize
\be
  \mathrm{Re}[F_{\text{AdS}}(-i\delta,\Lambda)]=\left.-\log |Z^{\text{D}}_{SU(2)}(a,\Lambda)|\right\vert_{a=-i\delta} \, ,
\ee
as a function of $\delta$, for various values of the coupling $\Lambda$. Only the origin $\delta=0$ corresponds to an $SU(2)$-preserving boundary condition, while for a nonzero $\delta$ the boundary $SU(2)$ current is broken to its Cartan. This can be seen from the non-conservation law \eqref{econstraintslinear}, with the 3d vector multiplet given by \eqref{eads}. This analogous to the adjoint Higgsing in the $\cN=2$ SYM theory in flat space. As discussed in the introduction, our expectations are that, at small AdS radius, the origin $\delta = 0$ will describe a stable $SU(2)$-preserving boundary condition, while, at finite $\Lambda L$, the preferred boundary condition chosen by $F_{\text{AdS}}$-maximization will be at finite $\delta$ and only preserve the Cartan of $SU(2)$.

We showcase only the main outcomes of our analysis, the reader interested in the detailed implementation can consult appendix \ref{app: numerics}.
As shown previously, the full supersymmetric partition function on AdS$_4$ for $SU(2)$ takes the form:
\be
Z^{\text{D}}_{SU(2)}(a,\Lambda )= \Lambda^{4 a^2}\frac{2 a}{\sinh(2 \pi a)}G(1+2 i a)G(1-2 i a) Z_{\text{Nekrasov}}(a,\Lambda)\, ,
\ee
where $a=m -i \delta$ parametrizes the complexified VEV of $\phi_1$, which without loss of generality we take to be in the $\tau_3$ direction: $\phi_1 = -\frac{a \tau_3}{2 \cosh(\eta)}$.
We will omit the explicit $L$ dependence in several expressions. It can be restored by rescaling $a \to a L$ and $\Lambda\to\Lambda L$.

The Nekrasov partition function admits a well known weak coupling expansion
\be
Z_{\text{Nekrasov}} = 1 + \sum_{k=1}^\infty I_k(a) \, \Lambda^{4k} \, .
\ee
The coefficients $I_k(a)$ of the instanton expansion can be computed to an arbitrarily large order by employing the algorithm of \cite{Nekrasov:2002qd,Alday:2009aq}. In practice, for numerical evaluations we will need to truncate the sum over $k$ to some $k_{\text{max}}$. In our study we will use up to $k_{\text{max}}=16$. The first few terms of the series are
\begin{equation}
\begin{aligned}\label{eq:firstterms}
    I_1(a)&=\frac{1}{2 \left(a^2+1\right)}\\
    I_2(a)&=\frac{8 a^2+33}{4 \left(a^2+1\right) \left(4 a^2+9\right)^2}\\
    I_3(a)&=\frac{8 a^4+99 a^2+366}{24 \left(a^2+1\right)  \left(4 a^2+9\right)^2\left(a^2+4\right)^2}\\
    I_4(a)&=\frac{256 a^8+7616 a^6+91276 a^4+521211 a^2+1109820}{384 \left(a^2+1\right)  \left(4 a^2+9\right)^2 \left(a^2+4\right)^3 \left(4 a^2+25\right)^2} \, .
\end{aligned}
\end{equation}
This expansion is clearly valid in the weak coupling region $\Lambda \ll 1$. We are however interested in the physics at strong coupling $\Lambda  \sim O(1)$. The convergence of the expansion at intermediate $\Lambda$ is a non-trivial matter.\footnote{For a recent discussion of the convergence properties of the instanton expansion, albeit in a different regime, see \cite{Arnaudo:2022ivo}.}
For generic (fixed) values of $a$, the coefficients $I_k(a)$ decay fast enough with $k$, making the truncation error for fixed $\Lambda$ bounded as a function of $a$. 

We can also assess the reliability of our truncation by looking at the singularities of $F_{\text{AdS}}$ as a function of $\delta$. Note that singularities of $F_{\text{AdS}}$ correspond to zeroes or poles of $Z$. We observe two types of singularities: spurious ones at the $\Lambda$-independent location $\delta =\frac{1}{2} \bZ$ for $|\delta|\geq 1$, that disappear as we increase $k_{\text{max}}$; and physical ones whose position depend on the value of $\Lambda$. The spurious singularities originate from the fact that
\begin{equation}
    Z_{1\text{-loop}}=\left.\frac{2a}{\sinh(2\pi a)}G(1+2i a)G(1-2i a)\right\vert_{a=-i\delta}
\end{equation}
has zeroes at $\delta = \frac{1}{2} \bZ$ for $|\delta|\geq1$.
We observe that these zeros are compensated by the poles of the Nekrasov partition function at $\delta = \frac{1}{2} \bZ$ for $|\delta|\geq1$. For a given $k_{\text{max}}$ only a finite number of such poles is present in $Z_{\text{Nekrasov}}$, namely the ones at $\delta = \frac{n}{2} $ with $2\leq|n| \leq k_{\text{max}}+1$, as visible from the first few terms in \eqref{eq:firstterms}. As  $k_{\text{max}}$ is increased, more poles appear in $Z_{\text{Nekrasov}}$ and the orders of the poles also increase. As a result, more spurious singularities of $F_{\text{AdS}}$ disappear. So for a given $k_{\text{max}}$ we can only trust the function $F_{\text{AdS}}$ in the region $|\delta|\leq\delta_{\text{max}}$ in which the spurious poles cancel. We observe that up to $16$ instantons, all spurious singularities for $|\delta|\leq4$ are canceled.

On the other hand, the physical, $\Lambda$-dependent singularities of $F_{\text{AdS}}$ come from zeroes in $Z_\text{Nekrasov}$. We observe that these zeroes gradually merge and disappear as $\Lambda$ increases, starting from the ones that are at smaller values of $|\delta|$, and moving to larger and larger values as $\Lambda$ increases. As a result, a larger portion of the $\delta$ axis becomes free from singularities as $\Lambda$ increases. Interestingly, we find evidence that the disappearing of zeroes is related to the emergence of new maxima in the AdS free-energy, see the discussion in appendix \ref{app: numerics} for more details.

In order to numerically investigate the emergence of new maxima at strong coupling, we first need to determine the region in the $(\delta,\Lambda)$ plane in which the truncated instanton expansion is reliable. We impose two constraints: first, the truncation error, estimated by the relative error between $k_{\text{max}}$ and $k_{\text{max}}-1$, must be small for generic $\delta$ and fixed $\Lambda$; second, the spurious singularities in $\delta$ must have disappeared in the region of $\delta$ that we consider. For $k_{\text{max}}=16$ we find the reliable region to be $0<\delta<2.466$ and $0<\Lambda<1.5$, for which the relative error in the free energy is smaller than $10^{-6}$. This allows us to reliably investigate the emergence of new maxima in part of the strong-coupling regime.

\begin{figure}[htbp]
  \centering

  %–––––––––––––––––– 1st row ––––––––––––––––––%
  \begin{subfigure}[b]{0.3\textwidth}
    \centering
    \includegraphics[width=\linewidth]{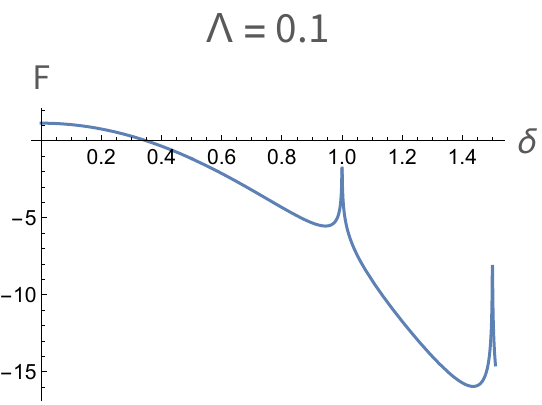}
    \caption{}\label{fig:F1}
  \end{subfigure}
  \begin{subfigure}[b]{0.3\textwidth}
    \centering
    \includegraphics[width=\linewidth]{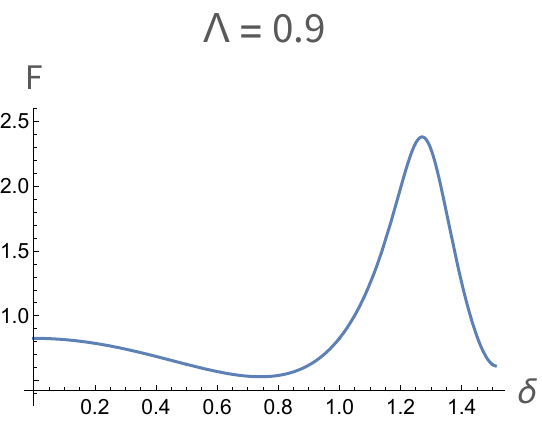}
    \caption{}\label{fig:F2}
  \end{subfigure}
  \begin{subfigure}[b]{0.3\textwidth}
    \centering
    \includegraphics[width=\linewidth]{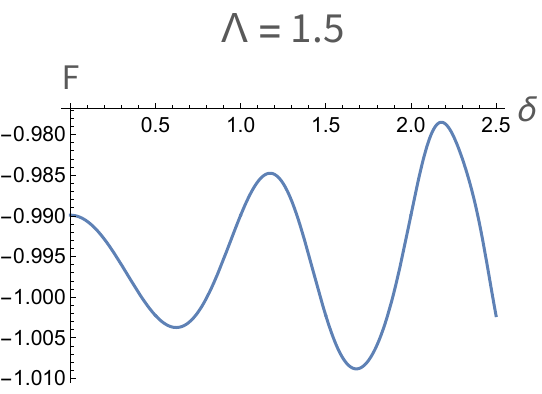}
    \caption{}\label{fig:F3}
  \end{subfigure}

  \caption{Free energy $F_{\text{AdS}}(-i\delta,\Lambda)$ for different values of $\Lambda$. (a) Weak coupling regime with single maximum at the origin $\delta=0$. Note that the points where the function shows steep growth are actual physical singularities where $\mathrm{Re}[F_{\text{AdS}}]\to+\infty$ and as such they do not give rise to any additional smooth local maximum outside of the origin. (b) At intermediate values of the coupling, a new smooth local maximum emerges at nonzero $\delta$, with a larger value of $\mathrm{Re}[F_{\text{AdS}
  }]$ compared to the one at origin, signaling the transition to a $SU(2)$-breaking boundary condition. (c) At even stronger coupling a second smooth local maximum with $\delta\neq 0$ emerges, larger than the first one. More and more of these maxima are expected to emerge as we increase $\Lambda L$, eventually accumulating and giving rise to the flat direction in the flat space limit.}
  \label{fig:F}
\end{figure}

\begin{figure}[htbp]
  \centering

  %–––––––––––––––––– 1st row ––––––––––––––––––%
  \begin{subfigure}[b]{0.4\textwidth}
    \centering
    \includegraphics[width=\linewidth]{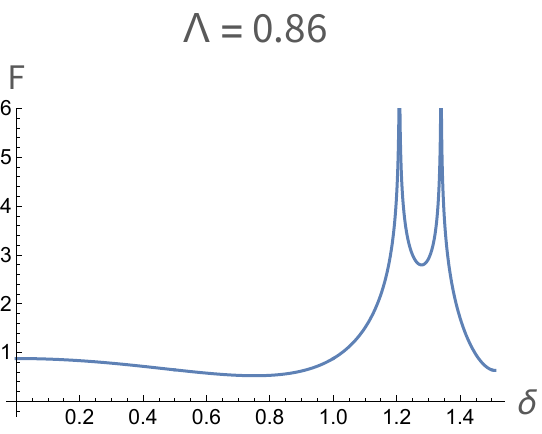}
    \caption{}\label{fig:FF1}
  \end{subfigure}\hspace{10mm}
  \begin{subfigure}[b]{0.4\textwidth}
    \centering
    \includegraphics[width=\linewidth]{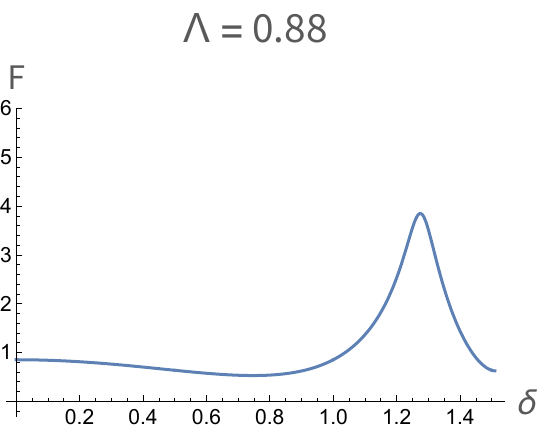}
    \caption{}\label{fig:FF2}
  \end{subfigure}

  \vspace{0.8em}

  %–––––––––––––––––– 2nd row ––––––––––––––––––%
  \begin{subfigure}[b]{0.4\textwidth}
    \centering
    \includegraphics[width=\linewidth]{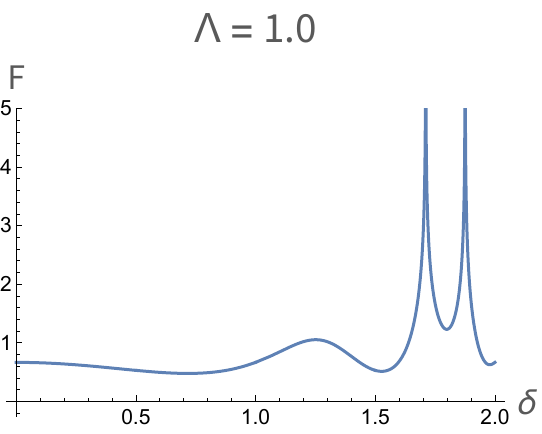}
    \caption{}\label{fig:FF3}
  \end{subfigure}\hspace{10mm}
  \begin{subfigure}[b]{0.4\textwidth}
    \centering
    \includegraphics[width=\linewidth]{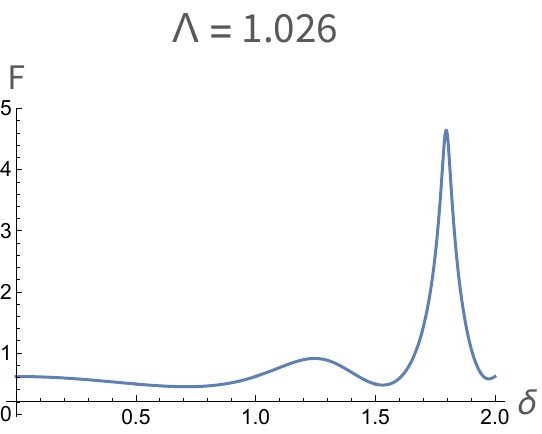}
    \caption{}\label{fig:FF4}
  \end{subfigure}

  \caption{New maxima in the free energy $F(-i\delta,\Lambda)$ emerge from merging singularities as $\Lambda$ increases. (a, b) A new leading maximum emerges at finite $\Lambda$ from the merging of two singularities. (c, d) A third maximum emerges from merging singularities. }
  \label{fig:FF}
\end{figure}

Figure \ref{fig:F} showcases the free-energy for different values of $\Lambda$. At weak coupling \ref{fig:F1} the origin of the Coulomb branch is a stable maximum, as predicted by the semiclassical analysis. This is the only maximum--defining a superconformal boundary condition with the full $SU(2)$ global symmetry-- for small enough $\Lambda$. Note that it is only a local and not a global maximum of the function, due to the existence of the physical singularities mentioned above, at which $\mathrm{Re}[F_{\text{AdS}}]\to +\infty$.

As we increase $\Lambda$ while remaining within our regime of validity, a second local maximum at $\delta\neq 0$ appears at $\Lambda\sim0.88$. This happens due to a merger of singularities (figures \ref{fig:FF1} and \ref{fig:FF2}). The value of $\mathrm{Re}[F_{\text{AdS}}]$ at this new local maximum is higher than the one at the origin. This is a new superconformal boundary condition where the bulk gauge symmetry is broken down to a $U(1)$ subgroup. Since this is the largest available local maximum, we take this to signal that the theory at $\Lambda > 0.88$ transitions to an adjoint Higgs phase, the one expected from the flat space limit.

By further increasing $\Lambda$, new leading maxima appear further away from the origin, while their overall height approaches zero, as shown in figure \ref{fig:F3} and figure \ref{fig:S} in the appendix. As none of the former maxima disappear, we expect that, in the flat space limit, a continuous Coulomb branch of vacua emerges from the merging of a dense set of discrete maxima of the AdS free energy.
This is also corroborated by the mechanism for the emergence of local maxima via merging of singularities of the instanton partition function, which requires the number of maxima to keep increasing as we approach the flat space limit.

Note that our analysis does not prove that the maximum at the origin becomes disallowed when the new maxima emerge, but it proves that a new $SU(2)$-breaking boundary condition becomes allowed, and it appears favored by the larger value of $\mathrm{Re}[F_{\text{AdS}}]$. Studying in more details the properties of the $SU(2)$ preserving boundary condition, such as the spectrum of its operators, it could be possible to find that some obstruction emerges at $\Lambda L\sim 1$ forcing one to adopt the new Higgsed boundary condition. Another related possibility is that the larger value of $\mathrm{Re}[F_{\text{AdS}}]$ in the new maximum signals the existence of a bulk instability with the $SU(2)$ preserving boundary condition, that triggers a transition to the broken one. It would be interesting to clarify this point, we leave it for future work.

Summarizing, and restoring the AdS length $L$ for clarity, our analysis leads to two main lessons:
\begin{enumerate}
    \item The local maximum of $\mathrm{Re}[F_{\text{AdS}}]$ is the $SU(2)$-preserving boundary condition $\delta=0$ at weak coupling $\Lambda L \ll 1$, but for $\Lambda L \sim 1$ larger maxima emerge with $\delta\neq 0$, breaking the boundary $SU(2)$ symmetry down to $U(1)$. This matches the expectation from the flat space limit, in which there is no vacuum preserving the full $SU(2)$ gauge group, and the gauge symmetry is broken to $U(1)$ by adjoint Higgsing. 
    \item The flat space Coulomb branch of vacua arises through the merging of local maxima of the AdS free energy $F_{\text{AdS}}$, which combine into an emergent flat direction for $\Lambda L \to +\infty$.
\end{enumerate}

\section{$F_{\text{AdS}}=-4 \pi i \mathcal{F}_{\text{AdS}}$}\label{sec:F=F}
In this section we provide evidence for an identification of the AdS-free energy $F_{\text{AdS}}$ with the curved space prepotential $\cF_{\text{AdS}}$. This is partly inspired by the results of \cite{Zan:2021ftf} in the context of AdS$_4$/CFT$_3$: that work found a relation between the $S^3$ partition function of the CFT$_3$ --which by the holographic duality can be also identified with the bulk AdS$_4$ partition function--  and the bulk prepotential in the supergravity theory. Besides the gravitational decoupling limit $M_{\text{Pl}}\to\infty$ of the result of  \cite{Zan:2021ftf}, the identification purely for QFT in AdS is suggested by three main observations:
\begin{enumerate}
    \item The AdS free energy $F_{\text{AdS}}$ obeys an analog of the Matone relation \cite{Matone:1995rx} \eqref{eq: matone} for the flat space prepotential, namely one can obtain the VEV of a gauge-invariant Coulomb branch operator by acting on it with a certain differential operator. 
    \item $F_{\text{AdS}}$ also reproduces the correct Seiberg-Witten prepotential in the flat space limit
    \be
    F_{\text{AdS}} \simeq -4 \pi i L^2 \cF_{\text{flat}}(a,\Lambda) + O(\log(L)) \, ,
    \ee
   where $\cF_{\text{flat}}$ reads, for the SU(2) theory\footnote{We use the conventions of \cite{Klemm:1995wp} with the following modifications: $a^2_{\text{here}} = 2 a^2_{\text{there}}$, $\Lambda_{\text{here}} = \frac{1}{2} \Lambda_{\text{there}}$.}
\be\label{eq:fsFprep}
    \cF_{\text{flat}}(a) = -\frac{i}{2\pi} a^2 \left(\log\left( \frac{\Lambda^2}{4 a^2} \right)+ 3 + \sum_{k=1}^\infty \cF_k \left(\frac{\Lambda}{a} \right)^{4k} \right)\, .
    \ee

    \item The maximization procedure can be recast as the positivity of the IR gauge coupling. Indeed at $\delta^*$
    \be
   \left. \frac{\partial^2\mathrm{Re}[F_{\text{AdS}}]}{\partial\delta^2} \right|_{\delta = \delta^*}\leq 0\,.
    \ee
    Using $\tau(a) = \cF_{\text{AdS}}''(a)$ and our proposed identification this becomes
    \be
    \left. \text{Im}\left[\tau(a) \right] \right|_{\delta = \delta^*} \geq 0 \, .
    \ee
\end{enumerate}
We now provide more details for the first two observations in the specific case of pure $SU(2)$ SYM. We believe that such a relationship holds more generally for $\mathcal{N}=2$ gauge theories in AdS background. 

\subsection{AdS Matone relation}
Let us define
\begin{align}
A_k=\text{Tr}[\phi^k]\,,\ \ \bar{A}_k=\text{Tr}[\bar{\phi}^k]\, .
\end{align}
Then the Matone relation in flat space reads
\be \label{eMatone}
u = -\langle A_2 \rangle = \pi i \left( \cF_{\text{flat}}(a) -\frac{1}{2} a \partial_a \, \cF_{\text{flat}}(a) \right) \, .
\ee
Since the prepotential \eqref{eq:fsFprep} is a function of the form $a^2 f(\Lambda/a)$, it satisfies the equation
\be
\left(2 - a \partial_a \right) \cF_{\text{flat}} = \Lambda \partial_\Lambda \cF_{\text{flat}} \, .
\ee
Recall that $\tau_{\text{UV}}$ appears in the combination $\Lambda = \mu e^{\pi i \tau_{\text{UV}}(\mu)/2}$.\footnote{A notational remark: the low energy coupling of the effective $U(1)$ gauge theory, which is obtained from the prepotential, is often denoted by $\tau$, hence to avoid confusion in this section we adopt the symbol $\tau_{\text{UV}}$ to denote the UV gauge coupling in the Lagrangian of the non-abelian gauge theory. }
This allows us to rewrite the Matone relation in the compact form
\be \label{eq: matonetau}
u = \partial_{\tau_{\text{UV}}} \cF_{\text{flat}} \, .
\ee

Moving to AdS, we can still define $u$ as the VEV of the operator $-\langle A_2 \rangle$, which for any choice of boundary condition that preserves the AdS isometries will be a constant. On the other hand, the $a$ dependence will stem from the choice of (modified) Dirichlet boundary condition. For general $a$, the AdS isometries are broken and there is a preferred foliation with $S^3$'s at fixed radial coordinate $\eta$. In this case we can think of $u$ as the expectation value at the center of AdS $\langle A_2(\eta= 0) \rangle$.

The relation \eqref{eq: matonetau} suggests that it should be possible to obtain $u =-\langle A_2(0) \rangle$ from the derivative with respect to the coupling in the Lagrangian of some quantity. To prove the AdS Matone relation we need to understand what this quantity is in AdS. Obtaining correlation functions of chiral primaries from derivatives with respect to the associated coupling is reminiscent of the results of \cite{Gerchkovitz:2016gxx} for Coulomb branch correlators on $S^4$. These have already been adapted in \cite{Bason:2023bin} to compute one-point functions of chiral operators on $HS^4$, and we will now discuss how to adapt them to AdS. 
Consider a generic deformation by the top component $\cC$ of a chiral operator of $\mathcal{R}$-charge $w$. The $\cN=2$ action is modified by
\be
S_{\cN=2} \rightarrow  S_{\cN=2} + \int_{\text{AdS}_4} \sqrt{g} \left( \lambda_k \cC_k + \bar{\lambda}_k \bar{\cC}_k \right) + \int_{\partial \text{AdS}_4} \sqrt{h} \left( \lambda_k \cC_{\partial, \, k} + \bar{\lambda}_k \bar{\cC}_{\partial, \, k} \right) \, ,
\ee
where the bulk Lagrangian densities have the following expression in terms of the components of the chiral multiplet described in appendix \ref{s4dmultiplets}
\begin{align}\label{ecalC}
\begin{split}
\mathcal C & = C -i ( w - 2 ) (\tau_3)^{AB} B_{AB} + 2 (w - 2)( w - 3) A\,,\\
\bar{\mathcal C} & = \bar{C} +i ( w - 2 ) (\tau_3)^{AB} \bar{B}_{AB} + 2 (w - 2)( w - 3) \bar{A}\,,
\end{split}
\end{align}
and $w$ denotes the conformal dimension of $A$.
The deformation transforms under $\cQ$ by a total derivative
\begin{align}\label{edeltaC}
\begin{split}
\delta\mathcal{C}&=-D_\mu\big(2i \bar{\epsilon}^A\sigma^\mu\Lambda_A+2s_2(w-2)(\tau_3)^{AB}\bar{\epsilon}_{(A}\bar{\sigma}^\mu\Psi_{B)}\big)\,,\\
\delta\bar{\mathcal{C}}&=-D_\mu\big(2i\epsilon^A\sigma^\mu\bar{\Lambda}_A-2s_1(w-2)(\tau_3)^{AB}\epsilon_{(A}\sigma^\mu\bar{\Psi}_{B)}\big)\,.
\end{split}
\end{align}
In the presence of a boundary, our deformation must be supplemented by the appropriate boundary term in order to preserve the $\cQ$ supercharge. Similarly to what we did in section \ref{s4dAdSsusy} for the SYM action, we define the boundary term on a cutoff surface at a large value of $\eta =\eta_0\gg 1$. The boundary terms read
\begin{align}
\begin{split}\label{eCperpads}
\mathcal{C}_\partial & =\coth\big(\tfrac{\eta_0}{2}\big)\bigg(-i (\tau_3)^{AB} B_{AB}+2 D_\perp A + 4 (w-2)A\\
&\ \ \ \ \ \ \ \ \ \ \ \ \ \ \ \ \ \ \ -(1+\coth^2\big(\tfrac{\eta_0}{2}\big)-6\coth\big(\tfrac{\eta_0}{2}\big)\coth(\eta_0)\big)A\bigg)~,\\
\bar{\mathcal{C}}_\partial & =\tanh\big(\tfrac{\eta_0}{2}\big)\bigg(i (\tau_3)^{AB} \bar{B}_{AB} +2 D_\perp \bar{A} + 4 (w-2)\bar{A}\\
&\ \ \ \ \ \ \ \ \ \ \ \ \ \ \ \ \ \ \ -(1+\tanh^2\big(\tfrac{\eta_0}{2}\big)-6\tanh\big(\tfrac{\eta_0}{2}\big)\coth(\eta_0)\big)\bar{A}\bigg)~.
\end{split}
\end{align}
In order to explicitly compute the derivative with respect to $\lambda_k$, we rewrite the perturbation $\cC$ as
\be
\cC = D_\mu V^\mu + \delta_{\cQ} F \, , \quad  \bar{\cC} = D_\mu \bar{V}^\mu + \delta_{\cQ} \bar{F}
\ee
where:
\bea
V^\mu &=  2B_{AB}\frac{\bar{\epsilon}^A\bar{\sigma}^\mu\epsilon^B}{\epsilon^C\epsilon_C}-2i\tau_{3,AB}\frac{\bar{\epsilon}^A\bar{\sigma}^\mu\epsilon^B}{\epsilon^C\epsilon_C}\bigg(s_2(2w-3)-s_1\frac{\bar{\epsilon}_C\bar{\epsilon}^C}{\epsilon^D\epsilon_D}\bigg)A-2\frac{\bar{\epsilon}_C\bar{\epsilon}^C}{\epsilon^D\epsilon_D}D^\mu A  \, , \\
\bar{V}^\mu &= -2\bar{B}_{AB}\frac{\epsilon^A \sigma^\mu\bar{\epsilon}^B}{\bar{\epsilon}_C\bar{\epsilon}^C}+2i\tau_{3,AB}\frac{\epsilon^A\sigma^\mu\bar{\epsilon}^B}{\bar{\epsilon}_C\bar{\epsilon}^C}\bigg(s_1(2w-3)-s_2\frac{\epsilon^C\epsilon_C}{\bar{\epsilon}_D\bar{\epsilon}^D}\bigg)A-2\frac{\bar{\epsilon}_C\bar{\epsilon}^C}{\epsilon^D\epsilon_D}D^\mu A   \, , \\
F &=  -\frac{\epsilon^A\Lambda_A}{\epsilon^C\epsilon_C}+i\frac{\tau_3^{AB}\epsilon_A\Psi_B}{\epsilon^C\epsilon_C}\bigg(s_2(w-3)-s_1\frac{\bar{\epsilon}_C\bar{\epsilon}^C}{\epsilon^D\epsilon_D}\bigg)-D_\mu\bigg(\frac{\bar{\epsilon}^A\bar{\sigma}^\mu\Psi_A}{\epsilon^C\epsilon_C}\bigg)     \, , \\
\bar{F} &=  -\frac{\bar{\epsilon}^A\bar{\Lambda}_A}{\bar{\epsilon}_C\bar{\epsilon}^C}+i\frac{\tau_3^{AB}\bar{\epsilon}_A\bar{\Psi}_B}{\bar{\epsilon}_C\bar{\epsilon}^C}\bigg(-s_1(w-3)+s_2\frac{\epsilon^C\epsilon_C}{\bar{\epsilon}_D\bar{\epsilon}^D}\bigg)+D_\mu\bigg(\frac{\epsilon^A\sigma^\mu\bar{\Psi}_A}{\bar{\epsilon}_C\bar{\epsilon}^C}\bigg)  \, .
\eea
A lengthy, but straightforward computation gives
\begin{align}
\begin{split}
&\int \mathcal{C}_k+\int_\partial\mathcal{C}_{\partial,k}=32\pi^2 A_k(0)+\delta_\mathcal{Q} \int F\,,\\
&\int \bar{\mathcal{C}}_k+\int_\partial\bar{\mathcal{C}}_{\partial,k}=\delta_\mathcal{Q} \int\bar{F}\,.
\end{split}
\end{align}
The $\cQ$-exact terms has zero VEV and we are left with the final result
\be\label{eq:derFads}
\partial_{\lambda_k} F_{\text{AdS}}\vert_{\lambda = 0} = 32 \pi^2 \langle A_k(0) \rangle \, ,
\ee
where the role of the north pole of the sphere in \cite{Gerchkovitz:2016gxx,Bason:2023bin} is played by the center of AdS$_4$. It might be confusing why the point $\eta=0$ plays a special role, given that AdS, unlike the hemisphere, is a homogeneous space. We stress that this is due to the fact, explained in section \ref{s4dAdSsusy}, that we can only manifestly preserve a subgroup of the (super)isometry, as is explicit by the choice of boundary terms on the cutoff surface $\eta=\eta_0$. Of course, if the boundary condition is chosen appropriately following the maximization procedure, the full spacetime (super)symmetry is restored and the VEV $\langle A_k \rangle$ is constant in AdS.

Let us specialize to $A=A_2$. In this case the associated chiral Lagrangian $\mathcal{C}_2$ gives precisely the SYM one in \eqref{eactionAdS4}, provided we specify 
\be
\lambda_2 = i \frac{\tau_{\text{UV}}}{8 \pi}~.
\ee
In this way we have 
\be \label{eAdSmatone}
\partial_{\tau_{\text{UV}}} F_{\text{AdS}} = 4 \pi i \langle A_2 (0) \rangle = -4\pi i \,u \, .
\ee
Comparison with \eqref{eq: matonetau} prompts the identification
\be\label{eq:FadsFprep}
F_{\text{AdS}} = -4 \pi i \cF_{\text{AdS}} \, .
\ee
We stress that the relation \eqref{eq:derFads} holds in general for Coulomb branch operators in any $\mathcal{N}=2$ theory. While for $SU(2)$ SYM the operators $A_k$ with $k\geq 3$ are not independent due to trace relations, some of them become independent for higher rank theories and this relation can be used to compute their VEVs. Moreover, both in higher rank theories and in $SU(2)$, one can also consider multiple derivatives, which by the same logic can be showed to give the VEV of multi-trace Coulomb branch operators. Taking the flat-space limit, these AdS relations allow to derive many generalizations of the Matone relation, both for higher rank and for multi-trace operators, relating their VEVs to appropriate derivatives of the flat-space prepotential with respect to UV couplings. It would be interesting to explore further these generalizations, both in AdS and in flat space.

\subsection{The flat space limit}
The identification \eqref{eq:FadsFprep} is also prompted by the flat space limit. We will work in a regime in which $L \to \infty$ while $a$ is kept finite. The relation between the flat-space prepotential and the flat-space limit of the supersymmetric partition function has been already pointed out on $S^4$ in \cite{Russo:2014nka,Hollowood:2015oma}, though in that case it is slightly obscured by the fact that it works at the level of the integrand of the localization matrix model. In the present context instead we are directly taking the limit of a partition function.
In this limit, the perturbative part of the partition function
\be
\log(Z^D_{SU(2)})_{\text{pert}} =  2 L^2 a^2 \log\left( \Lambda^2 \right) + \log\left( G(1 + 2 i a L) G(1 - 2 i a L) \frac{2 a L}{\sinh(2\pi a L)} \right)
\ee
becomes, via the large $z$ asymptotics $\log(G(1+z)) \sim \frac{1}{2} z^2 \log(z) -\frac{3}{4} z^2 + \hdots$ 
\be
\log(Z^D_{SU(2)})_{\text{pert}} \underset{L\to \infty}{\sim} 2 L^2 a^2\left( \log\left( \frac{\Lambda^2}{4 a^2} \right)+3\right) + O(\log(L)) \, ,
\ee
giving $4\pi i \, L^2$ times the perturbative part of the flat space prepotential \eqref{eq:fsFprep}. For the nonperturbative part \cite{Nekrasov:2002qd}, we use that in the small $\epsilon_1,\epsilon_2$ regime the instanton partition function approaches the nonperturbative part of the prepotential
\be
\log\left(Z_{\text{Nekrasov}}(a,\Lambda)\right) \underset{L\to \infty}{\sim} 4 
\pi i L^2 \cF_{\text{inst, flat}}(a) + ...~.
\ee
Putting the two pieces together
\be\label{ematchFF}
F(a,\Lambda) \underset{L \to \infty}{\sim} - 4 \pi i L^2 \cF_{\text{flat}}(a) + ...~,
\ee
consistently with our proposed identification \eqref{eq:FadsFprep}.

\section{Conclusions}
In this paper we have explored some aspects of $\cN=2$ SUSY QFTs on a background AdS space, with a special focus on the example of pure $SU(2)$ SYM. There are many interesting possible future directions:
\begin{itemize}
    \item \textbf{Higher rank/matter generalizations:} The study of the transitions between possible boundary conditions can be extended to theories with more general gauge groups, and/or with the inclusion of matter hypermultiplets \cite{Seiberg:1994aj}. The latter would require to include matter contributions in the localization calculation in AdS$_4$, though we still expect to find a match with the $HS^4$ result. Another interesting aspect would be to check the relation between the partition function and the prepotential in these more general theories. This relation suggests that in the higher rank case the bulk VEV of the various single trace chiral operators should be computed by the appropriate derivatives of the AdS prepotential. It would also be interesting to understand the mixing between multi-trace operators in AdS, which is related by our proposal to higher derivatives of the AdS prepotential \cite{Gerchkovitz:2016gxx}. 
    \item \textbf{Instantons in AdS:} We have seen that localized instantons are allowed by localization even in the presence of Dirichlet boundary conditions. The fact that these are singular configurations is presumably the reason why they violate the expectation that Dirichlet should forbid them \cite{Aharony:2012jf}. Their inclusion  is key to understand the strong coupling physics in AdS using localization. Besides the example studied here, the physics of instantons in AdS space remains largely unexplored. It would be instructive to understand how Dirichlet or Neumann boundary conditions on the conformal boundary lead to differences in the instanton spectrum, and the resulting effect on strong coupling physics.
    \item \textbf{Monopole and dyon points from $F_{\text{AdS}}$-maximization:} In flat space, new massless degrees of freedom emerge at the special loci $u=\pm \Lambda^2$ of the Coulomb branch, where either monopoles or dyons become massless. To make contact with these special points, it would be interesting to study $F_{\text{AdS}}$-maximization by allowing also a nonzero real part of $a$, and see what features emerge as we take the strong coupling limit. It is suggestive to note that the maximization condition $\mathrm{Re}[F_{\text{AdS}}']= 0$ is reminiscent of the condition $\mathcal{F}'(a)=0$ in Seiberg-Witten theory, describing the monopole point. It would also be interesting to see if there is any controllable AdS counterpart to the mechanism of confinement through monopole condensation, with a soft breaking to $\mathcal{N}=1$. However the soft breaking means that localization techniques used in this work would not be available and one would have to resort to some other method.
    \item \textbf{Inclusion of the $\theta$-angle:} In our work we have focused on the time-reversal invariant value $\theta=0$. For generic $\theta$ there is interesting new dynamics due the mixing between Dirichlet and Neumann boundary conditions, which has been studied in the context of the abelian BCFT with $\mathcal{N}=2$ supersymmetry in \cite{Herzog:2018lqz, KumarGupta:2019nay, Gupta:2020eev, Bason:2023bin}. In the asymptotically-free non-abelian case, we expect this to impact both the physics of AdS instantons and to give access to the analogue of the dyon point in the flat space limit.
    \item \textbf{Proving the conjectures about $F_{\text{AdS}}$:} It would be nice to connect $F_{\text{AdS}}$-maximization to the minimization of the bulk effective potential. The identification with the curved space prepotential is a first step in this direction, but the relation between the prepotential and the scalar potential in curved space \cite{lauria2020supergravity} is not straightforward enough to make such a connection evident. In turn a general proof of the relation $F_{\text{AdS}}= 4 \pi i \cF_{\text{AdS}}$ would be also interesting.
    \item \textbf{Monotonicity under boundary RG for QFTs in AdS:} Quantities that are extremized to determine the superconformal $\mathcal{R}$-symmetry often have definite monotonicity properties along RG flows, even when supersymmetry is absent, see e.g. \cite{Intriligator:2003jj, Komargodski:2011vj, Jafferis:2010un, Jafferis:2011zi, Klebanov:2011gs, Casini:2012ei, Gaiotto:2014gha, Casini:2018nym, Giombi:2020rmc}. This suggests that $F_{\text{AdS}}$ should decrease along RG flows, even for non-supersymmetric theories.
    More precisely, since for supersymmetric theories $F_{\text{AdS}}$ determines the boundary $\mathcal{R}$-symmetry, it should satisfy $F_{\text{AdS}}^{\text{UV}}>F_{\text{AdS}}^{\text{IR}}$ for boundary RG flows between two AdS-invariant boundary conditions with a fixed bulk theory. This conjecture can be seen as an extension many existing results about the monotonicity for BCFT and defect free energies \cite{Affleck:1991tk,Cuomo:2021rkm,Casini:2018nym,Casini:2022bsu,Casini:2023kyj,Jensen:2015swa,Wang:2020xkc,Wang:2021mdq,Giombi:2020rmc,Giombi:2021cnr} to the case of a massive bulk theory in AdS.
\end{itemize}
We leave these and other exciting questions for future work.
\paragraph{Acknowledgment} We thank M. Bertolini, G. Bonelli, F. De Cesare, A.G. Lezcano, S.S. Pufu and Xinyu Zhang for discussions. 
We especially thank Shota Komatsu for extensive discussion throughout this work and collaboration in related projects.
C.C. is supported by STFC grant
ST/X000761/1. ZJ is supported by the ERC-COG grant NP-QFT No. 864583 “Non-perturbative dynamics of quantum fields: from new deconfined phases of matter to quantum black holes” and by the MIUR-SIR grant RBSI1471GJ. LD and ZJ acknowledge support from the INFN “Iniziativa Specifica ST\&FI”. 

\appendix

\section{Conventions}\label{stensorconv}

In this appendix we describe various conventions and provide explicit coordinates which we refer to throughout the article. 

\subsection{General conventions}

\begin{enumerate}
	\item We work in Euclidean signature and use the following index notations to denote the different representations
\begin{equation}
\begin{aligned}[t]
&\mu,\ \nu,\ \hdots\ 4d\text{ curved-space indices}\tt{,}\notag\\
&i,\ j,\ \hdots\ 3d\text{ curved-space indices}\tt{,}\notag\\
&a,\ b,\ \hdots\ 4d\text{ tangent-space indices}\tt{,}\notag\\
&a',\ b',\ \hdots\ 3d\text{ tangent-space indices}\tt{,}\notag\\
&\alpha,\ \beta,\ \hdots\ \text{for }\psi_\alpha\in(\textbf{2},\textbf{1})_{\text{Spin}(4)}\text{,}\\
&\dot{\alpha},\ \dot{\beta},\ \hdots\ \text{for }\bar{\psi}^{\dot{\alpha}}\in(\textbf{1},\textbf{2})_{\text{Spin}(4)}\text{,}\\
&A,\ B,\ \hdots\ \text{for }\lambda_A\in\textbf{2}_{SU(2)_{\mathcal{R}}}\,,\\
&I,\ J,\ \hdots\ \text{adjoint gauge indices}\,.
\end{aligned}
\end{equation}
The spinors denoted with a bar $\bar{\psi}$ will always carry a dotted index as opposed to the unbarred ones. In Minkowski signature these spinors would be related to the unbarred ones by complex conjugation but in Euclidean signature they are independent.
\item We embed the tangent-space index $a'$ of the boundary inside the one of the bulk as $a=(a',4)$. Similarly for the space-time index we have $\mu=(i,\perp)=(i,\eta)$, where $\eta$ is the perpendicular coordinate described in \eqref{eadsmetrics}.
	\item The Pauli-matrices are taken to be
	\begin{align}
		\tau^1=\begin{pmatrix}
			0 & 1\\
			1 & 0
		\end{pmatrix}\,\ \ \ \tau^2=\begin{pmatrix}
			0 & -i\\
			i & 0
		\end{pmatrix}\,\ \ \ \tau^3=\begin{pmatrix}
			1 & 0\\
			0 & -1
		\end{pmatrix}\,.
	\end{align}
	\item All rank $n$ Levi-Civita tangent-space tensors are chosen such that $\varepsilon^{12\ldots n}= 1$. We define the space-time Levi-Civita by trading tangent-space and space-time indices with the vierbein.	
    \item The relevant spin and $\mathcal{R}$-symmetry tensors are defined below
    \begin{align}\label{e:gammamatrices}
		\begin{split}
			(\sigma^a)_{\alpha\dot{\beta}}=(-&i\vec{\tau},1)\,,\ \ \ \ (\bar{\sigma}^a)^{\dot{\alpha}\beta}=(i\vec{\tau},1)\,,\ \ \ \ (\gamma^{a'})_\alpha^{\ \beta}=\tau^{a'}\,,\ \ \ \ \vec{\tau}_A^{\ B}=\vec{\tau}\,,\\
			&\sigma^{ab}=\sigma^{[a}\bar{\sigma}^{b]}\,,\ \ \ \bar{\sigma}^{ab}=\bar{\sigma}^{[a}\sigma^{b]}\,,\ \ \ \gamma^{a'b'}=\gamma^{[a'}\gamma^{b']}\,.
		\end{split}
	\end{align}
	\item All spinorial and $SU(2)_R$-symmetry indices are raised and lowered with the $\varepsilon$-tensors as in the Wess \& Bagger notation \cite{Wess:1992cp}, namely
	\begin{equation}
		\varepsilon^{12}= \varepsilon_{21} = +1~.
	\end{equation}
	We employ the same notation for spinor bilinears as in \cite{Wess:1992cp}.

\item The covariant derivatives $D_\mu$ or $D_i$ include both space-time and gauge connections. To be explicit with our conventions, we display below the covariant derivative on spinors in some representation of the gauge group
\begin{align}
&D_\mu\chi_\alpha=\partial_\mu\chi_\alpha+\frac 14\omega_{\mu,ab}(\sigma^{ab}\chi)_\alpha-iA_\mu\chi_\alpha\,,\\
&D_\mu\bar{\chi}^{\dot{\alpha}}=\partial_\mu\bar{\chi}^{\dot{\alpha}}+\frac 14\omega_{\mu,ab}(\bar{\sigma}^{ab}\bar{\chi})^{\dot{\alpha}}-iA_\mu\bar{\chi}^{\dot{\alpha}}\,,
\end{align}
with
\begin{align}\label{espinconnection}
\omega_{\mu,ab}=-2\partial_{[\mu}e_{\nu][a}e^\nu_{b]}-e^\nu_{[a}e^\rho_{b]}\partial_{[\nu}e_{\rho]c}\,.
\end{align}
The 3d case is analogue. For adjoint fields we take $D_\mu\phi=\partial_\mu\phi-i[A_\mu,\phi]$.
\end{enumerate}

\subsection{Coordinate choices}\label{acoordinates}
We employ the following coordinates for AdS$_4$, setting for convenience the AdS radius to 1
\begin{align}\label{eadsmetrics}
ds^2=d\eta^2+\sinh(\eta)^2d\Omega_{S^3}^2\,,
\end{align}
with $\eta\in[0,+\infty)$. We then choose the following Hopf fibration for parametrizing the $S^3$ part of the metric
\begin{align}
z_1=\sin\bigg(\frac{\theta}{2}\bigg)e^{i\frac{\psi-\phi}{2}}\,,\ \ z_2=\cos\bigg(\frac{\theta}{2}\bigg)e^{i\frac{\psi+\phi}{2}}\,,\ \ \ \ |z_1|^2+|z_2|^2=1\,,
\end{align}
with $0\leq\theta\leq \pi\,\ 0\leq\phi\leq 2\pi\,\ 0\leq\psi\leq 4\pi$. With this the unit three-sphere has the following metric
\begin{align}
	d\Omega_{S^3}^2=\frac{1}{4}\big(d\theta^2+\sin^2(\theta)d\phi^2+(d\psi+\cos(\theta)d\phi)^2\big)\,.
\end{align}
A choice of vierbein is the following
\begin{align}
(e_{\text{AdS}_4})^a_{\ \mu}=\begin{pmatrix}
		0 & \frac 12 \sinh(\eta)\cos(\psi) & \frac 12 \sinh(\eta)\sin(\theta)\sin(\psi) & 0\\
		0 & -\frac 12 \sinh(\eta)\sin(\psi) & \frac 12 \sinh(\eta)\sin(\theta)\cos(\psi) & 0\\
		\frac 12 \sinh(\eta) & 0 & \frac 12 \sinh(\eta)\cos(\theta) & 0\\
		0 & 0 & 0 & 1
	\end{pmatrix}\,.
\end{align}

\newpage

\section{Supersymmetry multiplets}\label{asusytransf}
In this section we display the relevant multiplets, both in 4d and in 3d.

For the 4d case we follow the conventions in \cite{Bason:2023bin}. To obtain the transformations on the AdS$_4$ background we employ \eqref{Eq: background sugra values}-\eqref{e4dN2AdSHSKs} to fix the bosonic 4d supergravity content listed in table \ref{Tab: weyl multiplet}.

For 3d $\mathcal{N}=2$ supersymmetry we follow \cite{Closset:2012ru}, extending some transformations in the case of non-abelian vector fields. The bosonic 3d supergravity content is displayed in table \ref{tsugra3d}.
\begin{table}[h!]
\centering
\begin{tabular}{|c|c|c|}
\hline
\textbf{Bosons} & \textbf{Description} & \textbf{$\cR$-charge} \\ \hline
$g_{ij}$        & metric               & 0                     \\ \hline
$B_{ij}$    & 2-form gauge field   & 0                     \\ \hline
$C_i$         & photino              & 0                     \\ \hline
$A^{\cR}_i$   & $\cR$-gauge field    & 0                     \\ \hline
\end{tabular}
\caption{Bosonic content of the 3d $\mathcal{N}=2$ supergravity multiplet.}\label{tsugra3d}
\end{table}

\noindent
It is customary to redefine the photino and two-form as follows
\begin{align}
V^i=-i\varepsilon^{ijk}\partial_j C_k\,, \ \ \ \ \ H=\frac{i}{2}\varepsilon^{ijk}\partial_i B_{jk}\,.
\end{align}
On $S^3$ we have
\begin{align}
    H=\frac{i}{L}\,,\ \ \ V_i=A^{\mathcal{R}}_i=0\,,\ \ \ g_{ij}=g^{S^3}_{ij}\,,
\end{align}
and the Killing spinors satisfy
\begin{align}\label{e3dspherekse}
D_i\zeta=-\frac{i}{2}\gamma_i\zeta\,,\ \ \ D_i\tilde{\zeta}=-\frac{i}{2}\gamma_i\tilde{\zeta}\,.
\end{align}

\subsection{4d $\mathcal{N}=2$ multiplets}\label{s4dmultiplets}
\paragraph{Chiral multiplet}
{\allowdisplaybreaks
\begin{align}
&\delta A=- i\epsilon^A \Psi_A~,\notag\\
&\delta \Psi_A=2 \sigma^\mu \bar{\epsilon}_A D_\mu A + B_{AB}\epsilon^B + \tfrac{1}{2}\sigma^{\mu\nu}\epsilon_A G^-_{\mu\nu} + w \,\sigma^\mu D_\mu \bar{\epsilon}_A A~,\notag\\
& \delta B_{AB} = -2 i\bar{\epsilon}_{(A}\bar{\sigma}^\mu D_\mu \Psi_{B)} + 2i \epsilon_{(A}\Lambda_{B)} + i (1-w) D_\mu \bar{\epsilon}_{(A}\bar{\sigma}^\mu \Psi_{B)}~,\notag\\
& \delta G_{\mu\nu}^-= -\tfrac{i}{2}\bar{\epsilon}^A \bar{\sigma}^\rho \sigma_{\mu\nu} D_\rho \Psi_A -\tfrac{i}{2}\epsilon^A \sigma_{\mu\nu}\Lambda_A -\tfrac{i}{4}(1+w)D_\rho \bar{\epsilon}^{A}\bar{\sigma}^\rho \sigma_{\mu\nu}\Psi_{A}~,\notag\\
& \delta \Lambda_A = -\tfrac{1}{2} D_\rho G_{\mu\nu}^-\sigma^{\mu\nu}\sigma^\rho \bar{\epsilon}_A + D_\mu B_{AB}\sigma^\mu \bar{\epsilon}^B - C \epsilon_A\label{echiral}\\
&~~~~~~~~~+\tfrac{1}{2}(1+w)B_{AB}\sigma^\mu D_\mu\bar{\epsilon}^B + \tfrac{1}{4}(1-w)G^-_{\mu\nu}\sigma^{\mu\nu}\sigma^\rho D_\rho \bar{\epsilon}_A\notag\\
&~~~~~~~~~+ 4 D_\rho A \,\bar{T}^{\mu\nu}\sigma^\rho\bar{\sigma}_{\mu\nu}\bar{\epsilon}_A+4 w A D_\rho \bar{T}^{\mu\nu}\sigma^\rho \bar{\sigma}_{\mu\nu}\bar{\epsilon}_A~,\notag\\
& \delta C = -2 i \bar{\epsilon}^A \bar{\sigma}^\mu D_\mu \Lambda_A + 4 i \bar{T}^{\mu\nu}\bar{\epsilon}^A  \bar{\sigma}_{\mu\nu}\bar{\sigma}^\rho D_\rho \Psi_A \notag\\ 
& \,~~~~~~~ -i w D_\mu\bar{\epsilon}^A \bar{\sigma}^\mu \Lambda_A +4 i (w-1)D_\rho \bar{T}^{\mu\nu} \bar{\epsilon}^A \bar{\sigma}_{\mu\nu}\bar{\sigma}^\rho \Psi_A~.\notag
\end{align}
}
\paragraph{Anti-chiral multiplet}
{\allowdisplaybreaks
\begin{align}
&\delta \bar{A}= i\bar{\epsilon}^A \bar{\Psi}_A~,\notag\\
&\delta \bar{\Psi}_A=2 \bar{\sigma}^\mu \epsilon_A D_\mu \bar{A} + \bar{B}_{AB}\bar{\epsilon}^B +\tfrac{1}{2}\bar{\sigma}^{\mu\nu}\bar{\epsilon}_A G^+_{\mu\nu} + w \,\bar{\sigma}^\mu D_\mu \epsilon_A \bar{A}~,\notag\\
& \delta \bar{B}_{AB} = 2 i \epsilon_{(A}\sigma^\mu D_\mu \bar{\Psi}_{B)} + 2i \bar{\epsilon}_{(A}\bar{\Lambda}_{B)} - i (1-w) D_\mu \epsilon_{(A}\sigma^\mu \bar{\Psi}_{B)}~,\notag\\
& \delta G_{\mu\nu}^+= -\tfrac{i}{2}\epsilon^A \sigma^\rho \bar{\sigma}_{\mu\nu} D_\rho \bar{\Psi}_A -\tfrac{i}{2}\bar{\epsilon}^A \bar{\sigma}_{\mu\nu}\bar{\Lambda}_A +\tfrac{i}{4}(1+w)D_\rho \epsilon^{A}\sigma^\rho \bar{\sigma}_{\mu\nu}\bar{\Psi}_{A}~,\notag\\
& \delta \bar{\Lambda}_A = \tfrac{1}{2} D_\rho G_{\mu\nu}^+\bar{\sigma}^{\mu\nu}\bar{\sigma}^\rho \epsilon_A - D_\mu \bar{B}_{AB}\bar{\sigma}^\mu \epsilon^B + \bar{C}\bar{\epsilon}_A\label{eantichiral}\\
&~~~~~~~~~-\tfrac{1}{2}(1+w)\bar{B}_{AB}\bar{\sigma}^\mu D_\mu \epsilon^B - \tfrac{1}{4}(1-w)G^+_{\mu\nu}\bar{\sigma}^{\mu\nu}\bar{\sigma}^\rho D_\rho \epsilon_A\notag\\
&~~~~~~~~~+4 D_\rho \bar{A} \,T^{\mu\nu}\bar{\sigma}^\rho\sigma_{\mu\nu}\epsilon_A + 4 w \bar{A} D_\rho T^{\mu\nu}\bar{\sigma}^\rho \sigma_{\mu\nu}\epsilon_A~,\notag\\
& \delta \bar{C} = -2 i \epsilon^A \sigma^\mu D_\mu \bar{\Lambda}_A +4 i T^{\mu\nu} \epsilon^A  \sigma_{\mu\nu}\sigma^\rho D_\rho \bar{\Psi}_A\notag\\ 
& \,~~~~~~~ -i w D_\mu\epsilon^A \sigma^\mu \bar{\Lambda}_A +4 i (w-1) D_\rho T^{\mu\nu} \epsilon^A \sigma_{\mu\nu}\sigma^\rho \bar{\Psi}_A~.\notag
\end{align}
\paragraph{Vector multiplet}
{\allowdisplaybreaks
\begin{align}\label{e4dvector}
\begin{split}
&\delta\phi=-i\epsilon^A\lambda_A\,,\\
&\delta\bar{\phi}=i\bar{\epsilon}^A\bar{\lambda}_A\,,\\
&\delta\lambda_A=\frac{1}{2}\sigma^{\mu\nu}\epsilon_A(F_{\mu\nu}+8\bar{\phi} T_{\mu\nu})+2\sigma^\mu\bar{\epsilon}_A D_\mu\phi+\sigma^\mu D_\mu\bar{\epsilon}_A\phi+2i \epsilon_A[\phi,\bar{\phi}]+D_{AB}\epsilon^B\,,\\
&\delta\bar{\lambda}_A=\frac{1}{2}\bar{\sigma}^{\mu\nu}\bar{\epsilon}_A(F_{\mu\nu}+8\phi \bar{T}_{\mu\nu})+2\bar{\sigma}^\mu \epsilon_A D_\mu\bar{\phi}+\bar{\sigma}^\mu D_\mu\epsilon_A\bar{\phi}-2i\bar{\epsilon}_A[\phi,\bar{\phi}]+D_{AB}\bar{\epsilon}^B\,,\\
&\delta A_\mu=i\epsilon^A\sigma_\mu \bar{\lambda}_A-i\bar{\epsilon}^A\bar{\sigma}_\mu\lambda_A\,,\\
&\delta D_{AB}=-2i\bar{\epsilon}_{(A}\bar{\sigma}^\mu D_\mu\lambda_{B)}+2i\epsilon_{(A}\sigma^\mu D_\mu\bar{\lambda}_{B)}-4[\phi,\bar{\epsilon}_{(A}\bar{\lambda}_{B)}]+4[\bar{\phi},\epsilon_{(A}\lambda_{B)}]\,.
\end{split}
\end{align}
}
We can construct the following (anti-)chiral multiplet using the content of the vector multiplet, and for any given prepotential, i.e. any holomorphic function $\mathcal{F}(\phi)$ of the complex scalar in the vector multiplet:
\paragraph{Chiral multiplet from prepotential}
{\allowdisplaybreaks
\begin{align}
&A=\mathcal{F}(\phi)\,,\notag\\
&\Psi_{A\alpha}=\mathcal{F}_I\lambda^I_{A\alpha}\,,\notag\\
&B_{AB}=\mathcal{F}_ID^I_{AB}+\frac i2 \mathcal{F}_{IJ}\lambda_A^I\lambda^J_B\,,\notag\\
&G_{\mu\nu}^-=\mathcal{F}_I(F^{-I}_{\mu\nu}+8\bar{\phi}^IT_{\mu\nu})+2i\mathcal{F}_{IJ}\lambda^{IA}\sigma_{\mu\nu}\lambda_A^J\,,\notag\\
&\Lambda_{A\alpha}=\mathcal{F}_I(\slashed{D}\bar{\lambda}^I_A)_\alpha-i\mathcal{F}_I[\bar{\phi},\lambda_{A\alpha}]^I+\frac i4 \mathcal{F}_{IJ}(F^{-I}_{\mu\nu}+8\bar{\phi}^IT_{\mu\nu})(\sigma^{\mu\nu}\lambda^J_A)_\alpha\notag\\
&\ \ \ \ \ \ \ \ \ \!+\frac 12 \mathcal{F}_{IJ}D^I_{AB}\lambda^{JB}_\alpha-\frac{i}{12}\mathcal{F}_{IJK}\big(2(\lambda^I_A\lambda^{JB})\lambda^K_{B\alpha}-\lambda^I_{A\alpha}(\lambda^{JB}\lambda^K_B)\big)\,,\label{evectchiremb}\\
&C=-2\mathcal{F}_I\big(D^\mu D_\mu-\frac{\mathscr{R}}{6}+\frac{\tilde{M}}{2}\big)\bar{\phi}^I-8\mathcal{F}_I(F^{+I}_{\mu\nu}+8\phi^I\bar{T}_{\mu\nu})\bar{T}^{\mu\nu}\notag\\
&\ \ \ \ \ \ +\mathcal{F}_I[\bar{\lambda}^A,\bar{\lambda}_A]^I-2\mathcal{F}_I[[\bar{\phi},\phi],\bar{\phi}]^I+\frac 14 \mathcal{F}_{IJ} D^{IAB}D^J_{AB}\notag\\
&\ \ \ \ \ \ - \frac 12 \mathcal{F}_{IJ}(F^{-I}_{\mu\nu}+8\bar{\phi}^IT_{\mu\nu})(F^{-J\mu\nu}+8\bar{\phi}^JT^{\mu\nu})+ i \mathcal{F}_{IJ}\lambda^{IA}\slashed{D}\bar{\lambda}^J_A\notag\\
&\ \ \ \ \ \ -\mathcal{F}_{IJ}([\bar{\phi},\lambda^A]^I\lambda^J_A)+\frac i4 \mathcal{F}_{IJK}D^{ABI}\lambda^J_A\lambda^K_B\notag\\
&\ \ \ \ \ \ +\frac i8 \mathcal{F}_{IJK}(F^{-I}_{\mu\nu}+8\bar{\phi}^IT_{\mu\nu})\lambda^{JA}\sigma^{\mu\nu}\lambda^K_A+\frac{1}{24}\mathcal{F}_{IJKL}(\lambda^{IA}\lambda^{JB})(\lambda^K_A\lambda^L_B)\,.\notag
\end{align}
}
\paragraph{Anti-chiral multiplet from prepotential}
{\allowdisplaybreaks
\begin{align}
&\bar{A}=\bar{\mathcal{F}}(\bar{\phi})\,,\notag\\
&\bar{\Psi}^{A\dot{\alpha}}=\bar{\mathcal{F}}_I\bar{\lambda}^{IA\dot{\alpha}}\,,\notag\\
&\bar{B}^{AB}=\bar{\mathcal{F}}_ID^{IAB}-\frac i2 \bar{\mathcal{F}}_{IJ}\bar{\lambda}^{IA}\bar{\lambda}^{JB}\,,\notag\\
&G_{\mu\nu}^+=\bar{\mathcal{F}}_I(F^{+I}_{\mu\nu}+8\phi^I\bar{T}_{\mu\nu})-\frac{i}{8}\bar{\mathcal{F}}_{IJ}\bar{\lambda}^{IA}\bar{\sigma}_{\mu\nu}\bar{\lambda}_A^J\,,\notag\\
&\bar{\Lambda}^{A\dot{\alpha}}=-\bar{\mathcal{F}}_I(\bar{\slashed{D}}\lambda^I_A)^{\dot{\alpha}}+i\bar{\mathcal{F}}_I[\phi,\bar{\lambda}^{A\dot{\alpha}}]^I-\frac i4 \bar{\mathcal{F}}_{IJ}(F^{+I}_{\mu\nu}+8\phi^I\bar{T}_{\mu\nu})(\bar{\sigma}^{\mu\nu}\bar{\lambda}^{JA})^{\dot{\alpha}}\notag\\
&\ \ \ \ \ \ \ \ \ \!+\frac 12 \bar{\mathcal{F}}_{IJ}D^{IAB}\bar{\lambda}^{J\dot{\alpha}}_B+\frac{i}{12}\bar{\mathcal{F}}_{IJK}\big(2(\bar{\lambda}^{IA}\bar{\lambda}^{JB})\bar{\lambda}^{K\dot{\alpha}}_B-\bar{\lambda}^{IA\dot{\alpha}}(\bar{\lambda}^{JB}\bar{\lambda}^K_B)\big)\,,\label{evectantichiremb}\\
&\bar{C}=-2\bar{\mathcal{F}}_I\big(D^\mu D_\mu-\frac{\mathscr{R}}{6}+\frac{\tilde{M}}{2}\big)\phi^I-8\bar{\mathcal{F}}_I(F^{-I}_{\mu\nu}+8\bar{\phi}^I T_{\mu\nu})T^{\mu\nu}\notag\\
&\ \ \ \ \ \  -\bar{\mathcal{F}}_I[\lambda^A,\lambda_A]^I+2\bar{\mathcal{F}}_I[\phi,[\phi,\bar{\phi}]]^I+\frac 14 \bar{\mathcal{F}}_{IJ} D^{IAB}D^J_{AB}\notag\\
&\ \ \ \ \ \ - \frac 12 \bar{\mathcal{F}}_{IJ}(F^{+I}_{\mu\nu}+8\phi^I\bar{T}_{\mu\nu})(F^{+J\mu\nu}+8\phi^J\bar{T}^{\mu\nu})-i \bar{\mathcal{F}}_{IJ}\bar{\lambda}^{IA}\bar{\slashed{D}}\lambda^J_A\notag\\
&\ \ \ \ \ \ +\bar{\mathcal{F}}_{IJ}([\phi,\bar{\lambda}^A]^I\bar{\lambda}^J_A)-\frac i4 \bar{\mathcal{F}}_{IJK}D^{ABI}\bar{\lambda}^J_A\bar{\lambda}^K_B\notag\\
&\ \ \ \ \ \ +\frac i8 \bar{\mathcal{F}}_{IJK}(F^{+I}_{\mu\nu}+8\phi^I\bar{T}_{\mu\nu})\bar{\lambda}^{JA}\bar{\sigma}^{\mu\nu}\bar{\lambda}^K_A-\frac{1}{24}\bar{\mathcal{F}}_{IJKL}(\bar{\lambda}^{IA}\bar{\lambda}^{JB})(\bar{\lambda}^K_A\bar{\lambda}^L_B)\,.\notag
\end{align}
}

\subsection{3d $\mathcal{N}=2$ multiplets}\label{s3d}
\paragraph{Vector multiplet}
{\allowdisplaybreaks
\begin{align}
&\delta\sigma=-\zeta\tilde{\lambda}+\tilde{\zeta}\lambda\,,\notag\\
&\delta\lambda=\bigg(i(D+H\sigma)-\frac{i}{2}\varepsilon^{ijk}\gamma_k F_{ij}-i\gamma^i(D_i\sigma+iV_i\sigma)\bigg)\zeta\,,\notag\\
&\delta\tilde{\lambda}=\bigg(-i(D+H\sigma)-\frac{i}{2}\varepsilon^{ijk}\gamma_k F_{ij}+i\gamma^i(D_i\sigma-iV_i\sigma)\bigg)\tilde{\zeta}\,,\label{e3dvectorsusy}\\
&\delta A_i=-i(\zeta \gamma_i\tilde{\lambda}+\tilde{\zeta}\gamma_i\lambda)\,,\notag\\
&\delta D=D_i(\zeta\gamma^i \tilde{\lambda}-\tilde{\zeta}\gamma^i \lambda)-i V_i(\zeta\gamma^i\tilde{\lambda}+\tilde{\zeta}\gamma^i\lambda)-H(\zeta\tilde{\lambda}-\tilde{\zeta}\lambda)-[\sigma,\zeta\tilde{\lambda}+\tilde{\zeta}\lambda]\,.\notag
\end{align}
}
\paragraph{Linear multiplet}
{\allowdisplaybreaks
\begin{align}
&\delta J=i\zeta j+i\tilde{\zeta}\tilde{j}\,,\notag\\
&\delta j=\tilde{\zeta} K+i\gamma^i\tilde{\zeta}(j_i+i D_i J)+\tilde{\zeta}[\sigma,J]\,\notag\\
&\delta\tilde{j}=\zeta K-i\gamma^i\zeta(j_i-i D_i J)-\zeta[\sigma,J]\,,\label{e3dlinear}\\
&\delta j_i=i\varepsilon_{ijk}D^j(\zeta\gamma^k j-\tilde{\zeta}\gamma^k\tilde{j})+[\sigma,\zeta\gamma_i j+\tilde{\zeta}\gamma_i\tilde{j}]+i[J,\tilde{\zeta}\gamma_i\lambda-\zeta\gamma_i\tilde{\lambda}]\,,\notag\\
&\delta K=-i D_i(\zeta\gamma^i j+\tilde{\zeta}\gamma^i\tilde{j})+2iH(\zeta j+\tilde{\zeta}\tilde{j})-V_i(\zeta\gamma^i j+\tilde{\zeta}\gamma^i \tilde{j})+[\zeta\tilde{\lambda}+\tilde{\zeta}\lambda,J]~.\notag
\end{align}
}
This definition of the linear multiplet matches the one of \cite{Closset:2012ru} in the abelian case. We have written the transformations for a linear multiplet in the adjoint of the gauge group, but one can write the transformations for generic representations. The supersymmetry transformations close upon imposing
\be
\begin{split}\label{econstraintslinear}
&D_i j^i-[\sigma+i D,J]-i[\sigma,K]+[\tilde{\lambda},\tilde{j}]-[\lambda,j]=0\,,\\
&[F_{ij},J]=0\,.
\end{split}
\ee
A vector multiplet can be re-organized into a linear one in the following way
\begin{align}
\begin{split}\label{evectortolinear}
 &J=\sigma\,,\ \ \ j=i\tilde{\lambda}\,,\ \ \ \tilde{j}=-i\lambda\,,\\
&j_i=-\frac{i}{2}\varepsilon_{ijk}F^{jk}\,,\ \ \ K=D+H\sigma\,.   
\end{split}
\end{align}

\newpage

\section{From $\mathcal{N}=2$ on AdS$_4$ to $\mathcal{N}=2$ on $S^3$ }\label{app: kspinor}
In this section we discuss in more detail the supersymmetry algebra in 4d $\mathcal{N}=2$ theories and the specific case of AdS$_4$. We explain the relation to the 3d $\mathcal{N}=2$ superconformal algebra and its reduction to the 3d $\mathcal{N}=2$ supersymmetry algebra on $S^3$. In order to be self-contained we rewrite some of the formulas used in the main text. 

\subsection{4d $\mathcal{N}=2$ supersymmetry algebra and AdS$_4$}

The supersymmetry transformations are generated by
\begin{align}\label{appeQ}
    \mathcal{Q}=\epsilon^A Q_A+\bar{\epsilon}_A\bar{Q}^A\,,
\end{align}
where $\epsilon_A$ and $\bar{\epsilon}_A$ solve some Killing spinor equations, depending on the chosen background. It is possible to find the supersymmetry algebra by acting with $\mathcal{Q}^2$ on all the fields \cite{Hama:2012bg}. We only display, as reference, the action on the gaugino
\begin{align}
\mathcal{Q}^2\lambda_A=iv^\mu\partial_\mu \lambda_A+i[\Phi,\lambda_A]+\frac i4L_{ab}\sigma^{ab}\lambda_A+\bigg(\frac 32 w+\Theta\bigg)\lambda_A+\tilde{\Theta}_{AB}\lambda^B\,,
\end{align}
with
\begin{align}\label{eQsquaretransf}
\begin{split}
v^\mu&=2\bar{\epsilon}^A\bar{\sigma}^\mu\epsilon_A\,,\\
\Phi^I&=2i\bar{\phi}^I\epsilon^A\epsilon_A+2i\phi^I\bar{\epsilon}_A\bar{\epsilon}^A+v^\mu A_\mu^I\,,\\
L_{ab}&=D_{[a}v_{b]}+v^\mu\omega_{\mu,ab}\,,\\
w&=-\frac i2(\epsilon^A\sigma^\mu D_\mu\bar{\epsilon}_A+D_\mu\epsilon^A\sigma^\mu\bar{\epsilon}_A)\,,\\
\Theta &=-\frac{i}{4}(\epsilon^A\sigma^\mu D_\mu\bar{\epsilon}_A-D_\mu\epsilon^A\sigma^\mu\bar{\epsilon}_A)\,,\\
\tilde{\Theta}_{AB}&=-i\epsilon_{(A}\sigma^\mu D_\mu\bar{\epsilon}_{B)}+iD_\mu\epsilon_{(A}\sigma^\mu\bar{\epsilon}_{B)}\,,
\end{split}
\end{align}
where $\omega_{\mu,ab}$ is the spin connection \eqref{espinconnection} and we use $I$ as adjoint-index of the gauge group.

From this we can write, implicitly defining some bosonic symmetry operators, the following
\begin{align}\label{eQsquare}
\mathcal{Q}^2=-v^\mu P_\mu+\Phi^I G_I+L_{ab}M^{ab}+wD+\Theta \mathcal{R}_{U(1)}+\tilde{\Theta}_{AB}\mathcal{R}_{SU(2)}^{AB}\,.
\end{align}
Specifying everything to our AdS$_4$ background with \eqref{Eq: background sugra values} and imposing the Killing spinor equations \eqref{e4dN2AdSHSKs} the algebra reduces to
\begin{align}\label{eQsquareAdS}
\begin{split}
\mathcal{Q}^2\bigg|_{\text{AdS}}&=-v^\mu P_\mu+\Phi^I G_I+L_{ab}M^{ab}+\tilde{\Theta}_{AB}\mathcal{R}_{SU(2)}^{AB}\,,\\
\tilde{\Theta}_{AB}&=\tau_{3,AB}(\epsilon^C\epsilon_C-\bar{\epsilon}_C\bar{\epsilon}^C)\,,
\end{split}
\end{align}
so the $\mathcal{R}$-symmetry group becomes $U(1)_{\mathcal{R}}\subset SU(2)_{\mathcal{R}}$. In the following section we discuss the relation to the 3d $\mathcal{N}=2$ superconformal algebra. 

\subsection{AdS$_4$ $\mathcal{N}=2$ and 3d $\mathcal{N}=2$ superconformal algebra}\label{afrom4dto3dconf}
The 3d superconformal symmetry algebra is spannded by the following conformal Killing spinors \cite{Freedman:2013oja}
\begin{align}\label{e3dN2superconf}
	D_i\xi_A=-\frac{i}{2}\gamma_i\xi_A\,,\ \ \ D_i\tilde{\xi}_A=\frac{i}{2}\gamma_i\tilde{\xi}_A\,,
\end{align}
where $A$ is the $\mathbf{2}$ of $SO(2)\sim U(1)_{\mathcal{R}}$. We make the relation among the two algebras explicit using the coordinate described in appendix \ref{acoordinates}.

We solve the AdS$_4$ Killing spinor equations \eqref{e4dN2AdSHSKs} with the following ansatz
\begin{align}
	\begin{split}
		&\epsilon_A=e^{\frac \eta2}\epsilon_{+A}+e^{-\frac \eta2}\epsilon_{-A}\,,\\
		&\bar{\epsilon}_A=e^{\frac \eta2}\bar{\epsilon}_{+A}+e^{-\frac \eta2}\bar{\epsilon}_{-A}\,.\\
	\end{split}    
\end{align}
One can see that the equations for the perpendicular variable are immediately solved by
\begin{align}\label{eq:epsilonbrel}
	\epsilon_{\pm A}=\mp i\tau_{3 A}^{\ \ B}\sigma_4\bar{\epsilon}_{\pm B}\,.
\end{align}
With this the remaining independent equations are three and can be taken to be the following
\begin{align}\label{eintermediatestepkse}
    D_i \epsilon_A=\frac{i}{2}\tau_{3A}^{\ \ B}\sigma_i\bar{\epsilon}_B\,.
\end{align}
We then find
\begin{align}\label{eepsilonplusminus}
\begin{split}
D_i^{3d}\epsilon_{\pm A}=\frac{i}{2}\gamma_i^{3d}\epsilon_{\mp A}\,,
\end{split}
\end{align}
where with \q{3d} we here stress that we have removed all notions of the perpendicular coordinate, namely we define
\begin{align}
\begin{split}
\frac{(\gamma^i_{3d})_\alpha^{\ \beta}}{\sinh(\eta)}&=i(\sigma^i\bar{\sigma}^\perp)_\alpha^{\ \beta}=-i(\sigma^\perp\bar{\sigma}^i)^\beta_{\ \alpha}\,,\\
\sinh(\eta)(\gamma_i^{3d})_\alpha^{\ \beta}&=i(\sigma_i\bar{\sigma}^\perp)_\alpha^{\ \beta}=-i(\sigma^\perp\bar{\sigma}_i)^\beta_{\ \alpha}\,.
\end{split}
\end{align}
and we define
\begin{align}
    D_i\chi_\alpha=D_i^{3d}\chi_\alpha+\frac 12 \omega_{i, a4}(\sigma^{a4}\chi)_\alpha\,.
\end{align}
In all future 3d formulas, as well as the ones in the main text, the superscript \q{3d} is omitted.

With all these ingredients, by choosing
\begin{align}
	\xi_A=\epsilon_{+A}-\epsilon_{-A}\,,\ \ \ \tilde{\xi}_A=\epsilon_{+A}+\epsilon_{-A}\,,
\end{align}
we solve \eqref{e3dN2superconf}. The 4d $\mathcal{N}=2$ Killing spinors in function of the 3d $\mathcal{N}=2$ superconformal ones are
\begin{align}\label{exitoepsilon}
\begin{split}
&\epsilon_{A\alpha}=\sinh\left(\tfrac{\eta}{2}\right)\xi_{A\alpha}+\cosh\left(\tfrac{\eta}{2}\right)\tilde{\xi}_{A\alpha}\,,\\
&\bar{\epsilon}_{A\dot{\alpha}}=-i\tau_{3,A}^{\ \ B}\left(\cosh\left(\tfrac{\eta}{2}\right)\xi_A^\beta+\sinh\left(\tfrac{\eta}{2}\right)\tilde{\xi}_A^\beta\right)\sigma_{\perp,\beta\dot{\alpha}}\,.
\end{split}
\end{align}
We proceed to discuss the 3d $\mathcal{N}=2$ supersymmetry algebra which we explicitly preserve in the article.

\subsection{Preserving 3d $\mathcal{N}=2$ supersymmetry}\label{spreserving3dsusy}

We now need to link the $\xi_A$ with the $\zeta,\, \tilde{\zeta}$ 3d $\mathcal{N}=2$ Killing spinors on $S^3$ discussed in appendix \ref{s3d}. Their $U(1)_{\mathcal{R}}$-symmetry transformation is a phase transformation \cite{Closset:2012ru} which we define to be the following
\begin{align}
	\zeta'=e^{i\theta_{\mathcal{R}}}\zeta\,,\ \ \ \tilde{\zeta}'=e^{-i\theta_{\mathcal{R}}}\tilde{\zeta}'\,.
\end{align}
The $U(1)_{\mathcal{R}}\subset SU(2)_{\mathcal{R}}$ symmetry rotation acts on $\xi_A$ via the generator $\tau_3$ as displayed below
\begin{align}
	\xi_A '=\big(e^{i\theta_{\mathcal{R}} \tau_3}\big)_A^{\ B}\xi_B=\begin{pmatrix}
		e^{i\theta_{\mathcal{R}}}\xi_1\\
		e^{-i\theta_{\mathcal{R}}}\xi_2
	\end{pmatrix}_A\,.
\end{align}
Modulo a choice of normalization, we have
\begin{align}
	\zeta=\xi_1\,,\ \ \ \tilde{\zeta}=\xi_2\,.
\end{align}
From this relation it follows from \eqref{e3dN2superconf}
that $\zeta$ and $\tilde{\zeta}$ solve \eqref{e3dspherekse}.

Preserving 3d $\mathcal{N}=2$ supersymmetry means then specifying the AdS$_4$ $\mathcal{N}=2$ Killing spinors in \eqref{e4dN2AdSHSKs} to
\begin{align}\label{e4dto3drestriction}
	\begin{split}
		&\epsilon_{A\alpha}=\sinh\left(\tfrac{\eta}{2}\right)\xi_{A\alpha}=\sinh\left(\tfrac{\eta}{2}\right)\begin{pmatrix}
			\zeta_\alpha\\
			\tilde{\zeta}_\alpha
		\end{pmatrix}_A\,,\\
		&\bar{\epsilon}_{A\dot{\alpha}}=-i\tau_{3,A}^{\ \ B}\cosh\big(\frac{\eta}{2}\big)(\xi_B\sigma_{\perp})_\alpha=-i\tau_{3,A}^{\ \ B}\begin{pmatrix}
			(\zeta\sigma_{\perp})_{\dot{\alpha}}\\
			(\tilde{\zeta}\sigma_{\perp})_{\dot{\alpha}}
		\end{pmatrix}_B\,.
	\end{split}
\end{align}
With this we then have
\begin{align}\label{eQsquareS3}
\mathcal{Q}^2\bigg|_{3d}=-v^i P_i+\Phi^I G_I+L_{a'b'}M^{a'b'}+\tilde{\Theta}_{AB}\mathcal{R}_{SU(2)}^{AB}\,,
\end{align}
namely the algebra restricts to the $\mathcal{N}=2$ one on $S^3$. Referring to \eqref{eQsquaretransf} and \eqref{einsanedefinition}, we have
\begin{align}
    v^i=-2\tilde{\zeta}\gamma^i\zeta\,,\ \ \ \Phi=-4i\zeta\tilde{\zeta}(\phi_2+i\cosh(\eta)\phi_1)+v^iA_i\,,\ \ \  \tilde{\Theta}_{AB}=2\zeta\tilde{\zeta}\tau_{3,AB}\,.
\end{align}

\newpage

\section{1-loop determinant in AdS$_4$}\label{a1loopads}

In this appendix we show the computation of the 1-loop determinants, relevant for our localization computation, using the index theorem \cite{Atiyah:1967}. We have already anticipated that this theorem is applicable only on compact spaces but, due to the fact this theorem reproduced the result known via other methods in non-compact spaces \cite{Assel:2016pgi,GonzalezLezcano:2023cuh}, we are going to apply it to our case. 

\subsection{Gauge-fixing}\label{sgaugefixing}
Let us now comment on how to properly define a gauge-fixed path integral in order to run the localization argument. This requires the addition of ghost fields and of a BRST transformation with associated nilpotent charge $\mathcal{Q}_{\text{BRST}}$. The results for the locus, the classical contribution and the contribution of instantons obtained in section \ref{sAdSlocalization} are not affected by the gauge fixing. The presentation is slightly different depending on the choice of boundary conditions.

\paragraph{Dirichlet boundary conditions}
Indicating schematically all the fields as $\Psi$ and the ghosts as $\Psi_{\text{ghost}}$, the path integral we aim to compute is
\begin{align}\label{egaugefixedpathintegral1}
\int\mathcal{D}\Psi\mathcal{D}\Psi_{\text{ghost}} e^{-S[\Psi]-t\mathcal{Q}_{\text{loc}}V-\mathcal{Q}_{\text{BRST}}V_{\text{ghost}}}\,.
\end{align}
We use $c$ and $\bar{c}$ to denote the ghost and anti-ghost fields, $B$ to denote a Lagrange multiplier field, and $\Psi_{\text{loc}}$ to denote the configuration of the fields at the locus $\mathcal{Q}_{\text{loc}}V = 0$. We introduce a gauge-fixing functional $G[\Psi-\Psi_{\text{loc}}]$ compatible with localization, namely such that $G(0)=0$ in the locus. We take
\begin{align}
V_{\text{ghost}}=\text{Tr}\left[\bar{c}\big(-G[\Psi-\Psi^{\text{loc}}]+\tfrac \kappa 2 B\big)\right]\,,
\end{align}
with $\kappa$ an arbitrary constant, and
\begin{align}
&\mathcal{Q}_{\text{BRST}}c=\tfrac i2 [c,c]_+\,,\notag\\
&\mathcal{Q}_{\text{BRST}}\bar{c}=B\,,\\
&\mathcal{Q}_{\text{BRST}}B=0\,,\notag\\
&\mathcal{Q}_{\text{BRST}}\Psi=i [c,\Psi]_{\pm} \ \ \text{(except for}\ \mathcal{Q}_{\text{BRST}}A_\mu=D_\mu c\text{)}\,,\notag
\end{align}
where $[\cdot,\cdot]_\pm$ denotes commutator ($-$) or anti-commutator ($+$) depending on the Grassmann parity of the fields. The approach here is similar to the one of \cite{Pestun:2007rz}, but in the case of Dirichlet boundary conditions there is no need to add ghosts for ghosts for constant gauge transformations, since these are not allowed by the boundary condition. 

We can then upgrade the localization argument to include the gauge fixing by defining
\begin{align}\label{eQhatVhat}
\begin{split}
    \hat{\mathcal{Q}}&=\mathcal{Q}_{\text{loc}}+\mathcal{Q}_{\text{BRST}}\,,\\
    \hat{V}&=V+V_{\text{ghost}}\,,\\
    V&=\underset{\Psi_\text{odd}}{\sum}\Psi_\text{odd}(\hat{\mathcal{Q}}\,\Psi_\text{odd})^*\,,\\
\end{split}
\end{align}
where $\Psi_{\text{odd}}$ denotes all the fields with odd Grassmann parity, and by defining the following supersymmetry transformations for the ghosts
\begin{align}\label{esusyghosts}
\begin{split}
&\mathcal{Q}_{\text{loc}}c=\Phi^{\text{loc}}-\Phi\,,\\
&\mathcal{Q}_{\text{loc}}\bar{c}=0\,,\\
&\mathcal{Q}_{\text{loc}}B=iv^\mu\partial_\mu\bar{c}+iv^\mu A^{\text{loc}}_\mu \bar{c}\,,
\end{split}
\end{align}
where $v^\mu =2\bar{\epsilon}^A\bar{\sigma}^\mu\epsilon_A$. We can then write our original path integral in the following way
\begin{align}\label{egaugefixedpathintegral2}
\begin{split}
\int\mathcal{D}\Psi\mathcal{D}\Psi_{\text{ghost}} e^{-S[\Psi]-t\hat{\mathcal{Q}}\hat{V}}\,.
\end{split}
\end{align}
We can trade $\mathcal{Q}V$ for $\hat{\mathcal{Q}}\hat{V}$ because in this way we are just adding either a $\mathcal{Q}$-exact or a $\mathcal{Q}_{\text{BRST}}$-exact quantity. We have
\begin{align}
\int\hat{\mathcal{Q}}\hat{V}\big|_{\text{even}}=\int\mathcal{Q}V\big|_{\text{even}}+\int B\left(-G[\Psi-\Psi^{\text{loc}}]+\tfrac \kappa 2 B\right)\,,
\end{align}
which is positive semi-definite. Finding the saddle-point of $\int\mathcal{Q}V\big|_{\text{even}}$ as we did in section \ref{slocus} fixes $G[0]=0$ and we are thus left just with a gaussian integral in $B$.

\paragraph{Neumann boundary conditions}

The only difference, which plays a role in the 1-loop determinant computation below, is that in this case we have to take into account ghosts for ghosts $a_0,\,\bar{a}_0,\,c_0,\,\bar{c}_0\,,B_0$ as in \cite{Pestun:2007rz}. Indeed in this case the constant gauge transformations are actual gauge redundancies. They act on the ghost field as a shift $c\rightarrow c+k$. The approach we have described with Dirichlet remains qualitatively the same, with the following modifications. The BRST-charge in this case acts as follows
\begin{align}
\begin{split}
&\mathcal{Q}_{\text{BRST}}c=\tfrac i2 [c,c]_+ -a_0\,,\\
&\mathcal{Q}_{\text{BRST}}\bar{c}=B\,,\\
&\mathcal{Q}_{\text{BRST}}B=-i[a_0,\bar{c}]\,,\\
&\mathcal{Q}_{\text{BRST}}a_0=0\,,\\
&\mathcal{Q}_{\text{BRST}}\bar{a}_0=\bar{c}_0\,,\\
&\mathcal{Q}_{\text{BRST}}c_0=-i[a_0,B_0]\,,\\
&\mathcal{Q}_{\text{BRST}}\bar{c}_0=-i[a_0,\bar{a}_0]\,,\\
&\mathcal{Q}_{\text{BRST}}B_0=c_0\,,\\
&\mathcal{Q}_{\text{BRST}}\Psi=c^I G_I\Psi\ \ \ \text{(except for}\ \mathcal{Q}_{\text{BRST}}A_\mu=D_\mu c\text{)}\,,
\end{split}
\end{align}
and squares to $\mathcal{Q}_{\text{BRST}}^2=-a_0^I G_I$, with $I$ an adjoint index of the gauge group and $G_I$ the operator that implements a gauge transformation (so $c^IG_I\Psi=i[c,\Psi]_\pm$ for all fields except $c^IG_I A_\mu=D_\mu c$). We then need to define the following supersymmetry transformations
\begin{align}\label{esusyghosts2}
\begin{split}
\mathcal{Q}_{\text{loc}}a_0=\mathcal{Q}_{\text{loc}}\bar{a}_0=\mathcal{Q}_{\text{loc}}B_0=\mathcal{Q}_{\text{loc}}c_0=\mathcal{Q}_{\text{loc}}\bar{c}_0=0
\end{split}
\end{align}
and we take $V_{\text{ghost}}$ to be
\begin{align}
 V_{\text{ghost}}=\text{Tr}\left[\bar{c}\left(-G[\Psi-\Psi^{\text{loc}}]+\tfrac{\kappa_1}{2}B+iB_0\right)-c\left(\bar{a}_0-\tfrac{\kappa_2}{2} a_0\right)\right]\,.
\end{align}

\subsection{The index theorem}

Using $\hat{\mathcal{Q}}_{\text{loc}}$ in \eqref{eQhatVhat}, we can divide the fields in $\hat{\mathcal{Q}}_{\text{loc}}$-cohomology as $(X_{\text{even}},X_{\text{odd}})$, $\hat{\mathcal{Q}}_{\text{loc}}(X_{\text{even}},X_{\text{odd}})$ and possibly $\hat{\mathcal{Q}}_{\text{loc}}$-singlets.

We can then write the localization Lagrangian \eqref{eQhatVhat}, focusing on the quadratic part since it is the only one relevant for the localization computation, as
\begin{align}\label{eVandQV}
\begin{split}
    \hat{V}\big|_{\text{quad.}}&=
    \begin{pmatrix}
        \hat{\mathcal{Q}}X_{\text{even}} & X_{\text{odd}} 
    \end{pmatrix}
    \begin{pmatrix}
        D_{00} & D_{01}\\
        D_{10} & D_{11}
    \end{pmatrix}
    \begin{pmatrix}
        X_{\text{even}}\\
        \hat{\mathcal{Q}}X_{\text{odd}} 
    \end{pmatrix}\,,\\
    \hat{\mathcal{Q}}\hat{V}\big|_{\text{quad.}}&=
    \begin{pmatrix}
        X_{\text{even}} & \hat{\mathcal{Q}}X_{\text{odd}} 
    \end{pmatrix}
    \begin{pmatrix}
        -\hat{\mathcal{Q}}^2 & 0\\
        0 & 1
    \end{pmatrix}
    \begin{pmatrix}
        D_{00} & D_{01}\\
        D_{10} & D_{11}
    \end{pmatrix}
    \begin{pmatrix}
        X_{\text{even}}\\
        \hat{\mathcal{Q}}X_{\text{odd}} 
    \end{pmatrix}\\
    &\ \ \ \ -\begin{pmatrix}
        \hat{\mathcal{Q}}X_{\text{even}} & X_{\text{odd}} 
    \end{pmatrix}
    \begin{pmatrix}
        D_{00} & D_{01}\\
        D_{10} & D_{11}
    \end{pmatrix}
    \begin{pmatrix}
        1 & 0\\
        0 & \hat{\mathcal{Q}}^2
    \end{pmatrix}
    \begin{pmatrix}
        \hat{\mathcal{Q}}X_{\text{even}}\\
        X_{\text{odd}} 
    \end{pmatrix}\,.
\end{split}
\end{align}
With this we can write the 1-loop determinant
\begin{align}
    Z_{\text{1-loop}}=\sqrt{\frac{\text{det}(K_{\text{odd}})}{\text{det}(K_{\text{even}})}}=\sqrt{\frac{\text{det}_{X_{\text{odd}}}(\hat{\mathcal{Q}}^2)}{\text{det}_{X_{\text{even}}}(\hat{\mathcal{Q}}^2)}}=\sqrt{\frac{\text{det}_{\text{Coker.}D_{10}}(\hat{\mathcal{Q}}^2)}{\text{det}_{\text{Ker.}D_{10}}(\hat{\mathcal{Q}}^2)}}\,.
\end{align}
The first equality comes from $\text{det}(AB)=\text{det}(A)\text{det}(B)$. The second one is due to the fact that $D_{10}$ maps $X_{\text{even}}$ into $X_{\text{odd}}$ and $[\hat{Q},D_{10}]=0$. For this reason the ratio of the determinants receives contribution only from the unpaired eigenmodes, which by definition belong to the kernel and cokernel.

Calling $\lambda_n$ the $n^{\text{th}}$ eigenvalue, $k_n$ and $c_n$ its multiplicity in the kernel and cokernel respectively, we can write
\begin{align}\label{e1loopbydef}
    Z_{\text{1-loop}}=\sqrt{\underset{n}{\prod}\lambda_n^{c_n-k_n}}\,.
\end{align}
While $\lambda_n$ can be obtained by studying the action of $\hat{\mathcal{Q}}^2_{\text{loc}}$ on the $\hat{\mathcal{Q}}_{\text{loc}}$-complex, $c_n-k_n$ can be extracted from the index
\begin{align}\label{eindexbydef}
    \text{Ind}_{D_{10}}(t)=\text{Tr}_{\text{Ker.}D_{10}}(e^{-i\hat{\mathcal{Q}}^2 t})-\text{Tr}_{\text{Coker.}D_{10}}(e^{-i\hat{\mathcal{Q}}^2 t})=-\underset{n}{\sum}(c_n-k_n)e^{-i\lambda_n t}\,.
\end{align}
For elliptic or transversally elliptic operators, which we define below, there are considerable simplifications in computing the index.

As detailed in \cite{Pestun:2007rz, Assel:2016pgi}, one can construct the so called symbol by sending $\partial_\mu\rightarrow ip_\mu$ in $D_{10}$. If the symbol has kernel when all $p_\mu=0$, $D_{10}$ is said to be elliptic and the index is a finite Laurent series in $e^{-it}$. If the same is true but with a singled out $\hat{p}\in\{p_\mu\}$ that can remain nonzero, then the operator is said to be transversally elliptic along $\hat{p}_\mu$ and the index is in general a formal Laurent series in $e^{-it}$. In localization computations, as the following one, the operator we are computing the index of can be proved to be transversally elliptic.

When we have an operator that acts on functions on a manifold with at least a free $U(1)$ action generated by some Killing vector $\hat{v}$, and when this operator is elliptic or transversally elliptic along $\hat{v}$, one can use (a generalization of) the Atiyah-Bott fixed point formula \cite{Atiyah:1967}. This is precisely the case for $D_{10}$ and the $U(1)$ action defined by the Killing vector generated by $\hat{\mathcal{Q}}^2$. Letting $x_0$ be the fixed points of this $U(1)$ action, we have
\begin{align}\label{eindex}
\text{Ind}_{D_{10}}(t)=\underset{x_0}{\sum}\frac{\text{Tr}_{X_{\text{even}}}(e^{-i\hat{\mathcal{Q}}^2 t})\big|_{x_0}-\text{Tr}_{X_{\text{odd}}}(e^{-i\hat{\mathcal{Q}}^2 t})\big|_{x_0}}{\text{det}(\delta^\mu_\nu-\partial_\nu x'^\mu)}\,,
\end{align}
where $x'^\mu$ is the transformation of $x^\mu$ under the $U(1)$ action induced by $\hat{\mathcal{Q}}^2$. In this case the only fixed point is the one specified by $\eta=0$ in the coordinates displayed in appendix \ref{acoordinates}.

We now proceed in computing the index for the Dirichlet boundary conditions, we comment at the end the result for Neumann boundary conditions.

\subsection{Writing the index (Dirichlet boundary conditions)}

In the case of Dirichlet boundary conditions, where we do not have to introduce ghosts for ghosts as we explain in section \ref{sgaugefixing}, the $\hat{\mathcal{Q}}_{\text{loc}}$-complex is formed by
\begin{align}\label{eQcohomologyvec}
    X_{\text{even}}=(\phi_2,A_\mu)\,\ \ \ \ \ \ \ X_{\text{odd}}=(\Theta_{AB},\bar{c},c)\,,
\end{align}
with
\begin{align}
    \Theta_{AB}=\epsilon_{(A}\lambda_{B)}+\bar{\epsilon}_{(A}\bar{\lambda}_{B)}
\end{align}
plus the following $\hat{\mathcal{Q}}_{\text{loc}}$-singlet
\begin{align}
\Phi=2i\bar{\epsilon}_A\bar{\epsilon}^A\phi+2i\epsilon^A\epsilon_A \bar{\phi}-iv^\mu A_\mu\,.
\end{align}
$-v^\mu P_\mu\subset \hat{\mathcal{Q}}^2_{\text{loc}}$ \eqref{eQsquare} acts near the center of AdS$_4$ as
\begin{align}
i\mathcal{L}_v=i\bigg(x^1\frac{\partial}{\partial x^2}-x^2\frac{\partial}{\partial x^1}\bigg)+i\bigg(x^3\frac{\partial}{\partial x^4}-x^4\frac{\partial}{\partial x^3}\bigg)\,.
\end{align}
Defining
\begin{align}
    z_1=x_1+ix_2\,,\ \ \ \ \ \ z_2=x_3+i x_4
\end{align}
we can see that $i\mathcal{L}_v$ generates the following transformations
\begin{align}
    z_1'=e^{it}z_1\,,\ \ \ \ \ \ z_2'=e^{it}z_2\,.
\end{align}
With these coordinates it is easier to compute the terms appearing in \eqref{eindex}, namely
\begin{align}
    \text{det}(\delta^\mu_\nu-\partial_\nu x'^\mu)=|1-e^{it}|^4
\end{align}
and
\begin{align}\label{eactiononAofQ2}
\begin{split}
\begin{aligned}
e^{-i\hat{\mathcal{Q}}^2t}A_{z_1}&=e^{-i(-1+ia)t}A_{z_1}\,,\\
e^{-i\hat{\mathcal{Q}}^2t}A_{z_2}&=e^{-i(-1+ia)t}A_{z_2}\,,\\
e^{-i\hat{\mathcal{Q}}^2t}A_{\bar{z}_1}&=e^{-i(1+ia)t}A_{\bar{z}_1}\,,\\
e^{-i\hat{\mathcal{Q}}^2t}A_{\bar{z}_2}&=e^{-i(1+ia)t}A_{\bar{z}_2}\,,\\
e^{-i\hat{\mathcal{Q}}^2t}\phi_2&=e^{-i(ia)t}\phi_2\,,
\end{aligned}\ \ \ \ \ \ 
\begin{aligned}
e^{-i\hat{\mathcal{Q}}^2t}\Theta_{11}&=e^{-i(2+i a)t}\Theta_{11}\,,\\
e^{-i\hat{\mathcal{Q}}^2t}\Theta_{12}&=e^{-i(i a)t}\Theta_{12}\,,\\
e^{-i\hat{\mathcal{Q}}^2t}\Theta_{22}&=e^{-i(-2+i a)t}\Theta_{22}\,,\\
e^{-i\hat{\mathcal{Q}}^2t}\bar{c}&=e^{-i(i a)t}\bar{c}\,,\\
e^{-i\hat{\mathcal{Q}}^2t}c&=e^{-i(i a)t}c\,.
\end{aligned}
\end{split}
\end{align}
Putting everything together and using \eqref{eindex} we have
\begin{align}\label{eind}
\begin{split}
\text{Ind}_{D_{10}}(t)&=\text{Tr}_{\text{adj.}}(e^{at})\frac{\bigg(\big(2e^{it}+2e^{-it}+1\big)-\big(e^{2it}+e^{-2it}+3\big)\bigg)}{|1-e^{it}|^4}\\
&=-\text{Tr}_{\text{adj.}}(e^{at})\frac{1+e^{2it}}{(1-e^{it})^2}=-\underset{\alpha\in \Delta}{\sum}e^{\alpha\cdot a t}\times \frac{1+e^{2it}}{(1-e^{it})^2}\,.
\end{split}
\end{align}
$\alpha\cdot a$ is defined to be the eigenvalue of the adjoint matrix $a$ acting on the eigen-subspace corresponding to the root $\alpha\in\Delta$.

\subsection{Obtaining the products of the eigenvalues}
We have different possible answers of the 1-loop determinant depending on whether we expand the denominator, for each numerator's addend of \eqref{eind}, in powers of $e^{it}$ or $e^{-it}$. As it happens in these cases for localization computations in AdS \cite{Assel:2016pgi, GonzalezLezcano:2023cuh}, one needs to find the right answer via additional arguments after having examined all possibilities. 

For these different expansions one can rewrite \eqref{eind} in four different ways
\begin{align}\label{evariousindex}
    \begin{split}
        \text{Ind}_{D_{10}}^{\text{[a]}}(t)&=-\underset{\alpha\in \Delta}{\sum}e^{\alpha\cdot a t}\times \frac{1+e^{2it}}{(1-e^{it})^2}\,,\\
        \text{Ind}_{D_{10}}^{\text{[b]}}(t)&=-\underset{\alpha\in \Delta}{\sum}e^{\alpha\cdot a t}\times \frac{1+e^{-2it}}{(1-e^{-it})^2}\,,\\
        \text{Ind}_{D_{10}}^{\text{[c]}}(t)&=-\underset{\alpha\in \Delta}{\sum}e^{\alpha\cdot a t}\times \bigg(\frac{1}{(1-e^{it})^2}+\frac{1}{(1-e^{-it})^2}\bigg)\,,\\
        \text{Ind}_{D_{10}}^{\text{[d]}}(t)&=-\underset{\alpha\in \Delta}{\sum}e^{\alpha\cdot a t}\times \bigg(\frac{e^{2it}}{(1-e^{it})^2}+\frac{e^{-2it}}{(1-e^{-it})^2}\bigg)\,.
    \end{split}
\end{align}
For instance the resulting expansion of
\begin{align}\label{euglyind}
\begin{split}
    \text{Ind}_{D_{10}}^{\text{[a]}}(t)&=-\underset{\alpha\in \Delta}{\sum}e^{\alpha\cdot a t}(1+e^{2it})\bigg(\underset{k\ge 0}{\sum}e^{ikt}\bigg)^2\\
    &=-\underset{\alpha\in \Delta}{\sum}e^{-i(i\alpha\cdot a) t}\ -\ \underset{\alpha\in \Delta}{\sum}\underset{k>0}{\sum}2ke^{-i(i\alpha\cdot a-k) t}\,,
\end{split}
\end{align}
using \eqref{e1loopbydef}-\eqref{eindexbydef}, leads to
\begin{align}\label{eugly1loop1}
    Z_{\text{1-loop}}^{[a]}=\underset{\alpha\in\Delta^+}{\prod} \alpha\cdot a \ \underset{k>0}{\prod}\big(k^2+(\alpha\cdot a)^2\big)^k\,,
\end{align}
with $\Delta^+$ being the set of positive roots. Coincidentally Ind$_{D_{10}}^{\text{[b]}}$ gives
\begin{align}\label{eugly1loop2}
    Z_{\text{1-loop}}^{[b]}=Z_{\text{1-loop}}^{[a]}\,,
\end{align}
but the other two choices give different answers
\begin{align}\label{e1loops}
    \begin{split}
        Z_{\text{1-loop}}^{[c]}&=\underset{\alpha\in\Delta^+}{\prod}(\alpha\cdot a)^2\ \underset{k> 0}{\prod}(k+i \alpha\cdot a)^{k+1}(k-i \alpha\cdot a)^{k+1}\,,\\
        Z_{\text{1-loop}}^{[d]}&=\underset{\alpha\in\Delta^+}{\prod}\underset{k> 0}{\prod}(k+i \alpha\cdot a)^{k-1}(k-i \alpha\cdot a)^{k-1}\,.
    \end{split}
\end{align}
We now discuss the regularization of these products.

\subsubsection{Regularizing the products}\label{aregularizing}

In order to regularize the 1-loop infinite products that we have found, we can use these two expressions for the $\Gamma$-function and Barnes $G$-function
\begin{align}
    \begin{split}
        \Gamma(z)&=\frac{e^{-\gamma_E z}}{z}\underset{k>0}{\prod}\bigg(1+\frac{z}{k}\bigg)^{-1}e^{\frac{z}{k}}\,,\\
        G(z)&=(2\pi)^{\frac{z}{2}}e^{-\frac{1+\gamma_E z^2+z}{2}}\underset{k>0}{\prod}\bigg(1+\frac{z}{k}\bigg)^k e^{-z+\frac{z^2}{k}}\,,
    \end{split}
\end{align}
where $\gamma_E$ the Euler-Mascheroni constant. With these equations we can write the following formal products
\begin{align}\label{eregularproducts}
    \begin{split}
        \underset{k>0}{\prod}(k+z)&=\underset{k>0}{\prod}ke^{\frac{z}{k}}\ \times\ \frac{e^{-\gamma_E z}}{\Gamma(z+1)}\,,\\
        \underset{k>0}{\prod}(k+z)^k&=\underset{k>0}{\prod}k^ke^{z-\frac{z^2}{2k}}\ \times\ G(1+z)\frac{e^{\frac{1+\gamma_E z^2+z}{2}}}{(2\pi)^{\frac{z}{2}}}\,.
    \end{split}
\end{align}
Using the expression above, and keeping for now all infinite constants, one gets the following results for 1-loop determinants in \eqref{eugly1loop1}-\eqref{eugly1loop2}-\eqref{e1loops}
\begin{align}
    \begin{split}
        Z_{\text{1-loop}}^{[a]}&=Z_{\text{1-loop}}^{[b]}=\big(e\underset{k>0}{\prod}k^{2k}\big)^{\text{rk}(G)}\ \times\ \underset{\alpha\in\Delta^+}{\prod}\underset{k>0}{\prod}e^{\frac{(\alpha\cdot a)^2}{k}}\\
        &\ \ \ \ \ \ \ \ \ \ \ \ \ \ \ \ \underset{\alpha\in\Delta^+}{\prod}\alpha\cdot a \ G(1+i\alpha\cdot a)G(1-i\alpha\cdot a)e^{-(\gamma_E \alpha\cdot a)^2}\,,\\
        Z_{\text{1-loop}}^{[c]}&=\big(e\underset{k>0}{\prod}(-1)^{1+k}k^{2k+2}\big)^{\text{rk}(G)}\ \times\ \underset{\alpha\in\Delta^+}{\prod}\underset{k>0}{\prod}e^{\frac{(\alpha\cdot a)^2}{k}}\\
        &\ \ \ \ \ \underset{\alpha\in\Delta^+}{\prod}\alpha\cdot a \ \sinh(\pi \alpha\cdot a)G(1+i\alpha\cdot a)G(1-i\alpha\cdot a)e^{-\gamma_E (\alpha\cdot a)^2}\,,\\
        Z_{\text{1-loop}}^{[d]}&=\big(e\underset{k>0}{\prod}(-1)^{1+k}k^{2k+2}\big)^{\text{rk}(G)}\ \times\ \underset{\alpha\in\Delta^+}{\prod}\underset{k>0}{\prod}e^{\frac{(\alpha\cdot a)^2}{k}}\\
        &\ \ \ \ \ \underset{\alpha\in\Delta^+}{\prod} \frac{\alpha\cdot a}{\sinh(\pi \alpha\cdot a)}G(1+i\alpha\cdot a)G(1-i\alpha\cdot a)e^{-\gamma_E (\alpha\cdot a)^2}\,,
    \end{split}
\end{align}
with rk$(G)$ being the rank of the gauge group. The $a$-independent quantities can be safely removed
\begin{align}
\begin{split}\label{eadsreg1loopdirichlet}
Z_{\text{1-loop}}^{[a]}&=Z_{\text{1-loop}}^{[b]}=\underset{\alpha\in\Delta^+}{\prod}\underset{k>0}{\prod}e^{\frac{(\alpha\cdot a)^2}{k}}\ \times\ \underset{\alpha\in\Delta^+}{\prod}\alpha\cdot a\ G(1+i\alpha\cdot a)G(1-i\alpha\cdot a)e^{-\gamma_E(\alpha\cdot a)^2}\,,\\
Z_{\text{1-loop}}^{[c]}&=\underset{\alpha\in\Delta^+}{\prod}\underset{k>0}{\prod}e^{\frac{(\alpha\cdot a)^2}{k}}\ \times\ \underset{\alpha\in\Delta^+}{\prod}\alpha\cdot a\ \sinh(\pi \alpha\cdot a)G(1+i\alpha\cdot a)G(1-i\alpha\cdot a)e^{-\gamma_E (\alpha\cdot a)^2}\,,\\
Z_{\text{1-loop}}^{[d]}&=\underset{\alpha\in\Delta^+}{\prod}\underset{k>0}{\prod}e^{\frac{(\alpha\cdot a)^2}{k}}\ \times\ \underset{\alpha\in\Delta^+}{\prod} \frac{\alpha\cdot a}{\sinh(\pi \alpha\cdot a)}G(1+i\alpha\cdot a)G(1-i\alpha\cdot a)e^{-\gamma_E (\alpha\cdot a)^2}\,.
\end{split}
\end{align}
We remain with an infinite quadratic part $\sim e^{\underset{k}{\sum}\tfrac{\alpha\cdot a}{k}}$ which can be absorbed, together with the finite piece coming from $\sim e^{-\gamma_E(\alpha\cdot a)^2}$, into the renormalization of $g^2$. 

\subsection{Vector multiplet with Dirichlet boundary condition}

The partition function we are working with is invariant under $\phi_1\rightarrow-\phi_1$. For this reason we rule out [a] and [b], since with these 1-loop determinants the partition function is not invariant under $a$ to $-a$. We also do not expect the 1-loop contribution to be zero at the origin of the Coulomb branch, because this would give a vanishing partition function in that point. We thus rule out [c]. We then expect the right partition function for Dirichlet boundary conditions to be [d]. This coincides with the Dirichlet 1-loop contribution of $HS^4$ \cite{Bason:2023bin}.

\subsection{Vector multiplet with Neumann boundary condition}

We are not going to present the full argument, but the $\hat{\mathcal{Q}}_{\text{loc}}$-cohomology is similar to the one in \cite{Pestun:2007rz} and one needs to include also the ghost for ghosts, as explained in section \ref{sgaugefixing}. Again from \cite{Pestun:2007rz} it can be seen that their introductions contributes to a +2 in the index in \eqref{eind}. Repeating thus the argument of the index we find again four possible 1-loop results which are equivalent to the ones showed in \eqref{eadsreg1loopdirichlet} apart for the multiplication by a factor of $\underset{\alpha\in\Delta^+}{\prod}\frac{1}{(\alpha\cdot a)^2}$. The only expressions invariant under $a\rightarrow -a$ are again [c] and [d]. The 1-loop contribution [d] becomes singular in $a=0$, after the multiplication of $\underset{\alpha\in\Delta^+}{\prod}\frac{1}{(\alpha\cdot a)^2}$. We interpret this as a signal to discard this partition function and we keep [c] multiplied$\underset{\alpha\in\Delta^+}{\prod}\frac{1}{(\alpha\cdot a)^2}$.\footnote{This is more a signal, as opposed to a strict argument, since the partition function remains finite. Indeed with Neumann boundary conditions we still possess an integration over the gauge algebra matrices $\int_{\mathfrak{g}}da$ and its reduction to the Cartan is $\int_{\mathfrak{h}\subset\mathfrak{g}}da \underset{\alpha\in\Delta^+}{\prod}(\alpha\cdot a)^2$. The partition function would thus turn out to be finite both for [c] and [d].} With this the partition function result matches again the one on $HS^4$ \cite{Gava:2016oep,Dedushenko:2018tgx}.

\section{Details on the numerics}\label{app: numerics}
Here we provide more details on the numerics. In figure \ref{fig:F1}, it is clear that the origin $\delta=0$ is the only maximum of the free energy at small $\Lambda$. The numerics hint that the $SU(2)$ symmetric massless boundary condition is the only superconformal boundary condition in a finite range of $\Lambda$ above 0. As we increase $\Lambda$, the maximum at the origin remains, but $F(a,\Lambda)$ changes dramatically. In particular, the singularities move, see figure \ref{fig:FF}. As we have mentioned earlier in section \ref{snewmaxima}, there are $\Lambda$-dependent singularities in $F(-i\delta,\Lambda)$ which we conjecture to disappear in order to be consistent with the Seiberg-Witten theory in the flat space limit. Note that all the singularities present in figure \ref{fig:F} and \ref{fig:FF} are the $\Lambda$-dependent singularities, as the $\Lambda$-independent singularities at the half-integer value of $\delta$ have already been canceled by the 16-instanton terms we have included. We stress that the singularities in figure \ref{fig:F1} are not at half-integer values, indeed these singularities move as can be seen from figure \ref{fig:FF1}.

At a critical value near $\Lambda\sim0.88$, a bigger maximum at $\delta^*\sim1.27$ emerges, which is presented in figure \ref{fig:FF}. From figure \ref{fig:FF1} to figure \ref{fig:FF2}, one observes that two singularities merge and generate the new maximum when $\Lambda$ increases. Within the reliable region of our numerics, this second maximum coexists with the maximum at the origin for $\Lambda>0.88$. A similar mechanism is responsible for a third maximum emerging at a higher $\Lambda\sim1.026$, see figure \ref{fig:FF3} and \ref{fig:FF4}.

Within our control of the numerics, we do not observe the disappearance of the second maximum in figure \ref{fig:F2} or the third maximum in figure \ref{fig:F3}. We conjecture that they exist all the way to the flat space limit $\Lambda\to\infty$. We also conjecture that there are infinitely many such new maxima as $\Lambda\to\infty$. We expect this to be true because there seem to be infinitely many zeros in the full $Z_\text{inst}$, and they need to disappear in the flat space limit for the prepotential to be analytic in $a$. The free energy has the form $F(a)=-\frac{1}{2}\log f(a)$, where $f(a)=|Z^{\text{D}}_{\text{pure SYM}}(a,\Lambda)|^2$. For $F(a)$ to be real, $f(a)$ must be positive as it is in our case. Thus the zeros in $f(a)$ must have even orders. Then $F(a)$ must have positive first derivatives near the left of the singularity and negative second derivatives near the right. When two such singularities merge continuously and leave a continuous function, there is at least one point in a small vicinity of the merger that the first derivative vanishes and the second derivative is negative, hence a local maximum. Our conjecture states that such maxima remain in the large $\Lambda$ limit.

The second maximum near $\delta^*\sim1.27$ is very well within our control of numerics. Near the second maximum, the reliable $\Lambda$ can be up to $\Lambda\sim2$. Now we proceed to analyze the second maximum numerically, focusing on its behavior as $\Lambda$ increases. From the log-log plot \ref{fig:location} we observe that the location of the second maximum $\delta^*$ decays in a power law with $\Lambda$: $\delta^*\sim1.25\Lambda^{-0.16}$. From the log plot \ref{fig:peak}, we observe that the difference between the second maximum and the maximum at the origin decays exponentially with $\Lambda$, the best fit being $F(\delta^*,\Lambda)-F(0,\Lambda)\sim1772e^{-8.56\Lambda}$. The numerics suggest that as $\Lambda$ increases, the new maximum tends to be lower and moves toward the origin. We observe a similar power law and exponential scaling for the third maximum in a reliable range of $\Lambda$. Based on these numerics, we form the last conjecture in section \ref{snewmaxima}: as $\Lambda$ increases, an infinite set of discrete new maxima condense and become a dense set on the imaginary $a$ axis and connect to the Coulomb branch of vacua in the flat space limit, reproducing predictions of the Seiberg-Witten theory.

\begin{figure}[htbp]
  \centering
  \begin{subfigure}[b]{0.48\textwidth}
    \centering
    \includegraphics[width=\textwidth]{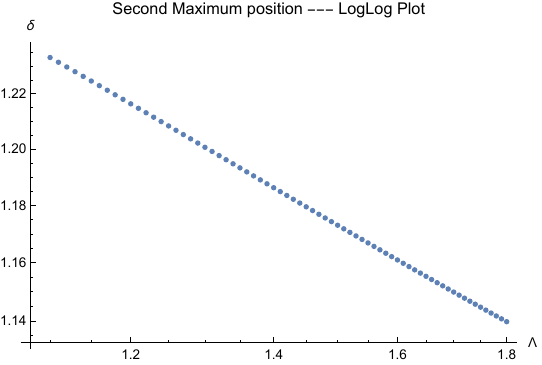}
    \caption{The location of the second maximum on the real $\delta$ axis as a function of $\Lambda$.}
    \label{fig:location}
  \end{subfigure}
  \quad
  \begin{subfigure}[b]{0.48\textwidth}
    \centering
    \includegraphics[width=\textwidth]{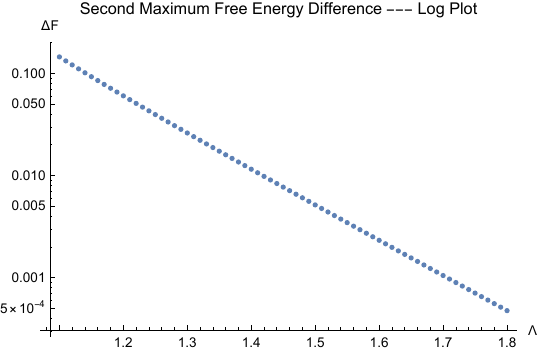}
    \caption{The difference between the free energy of the second and the first maximum as a function of $\Lambda$.}
    \label{fig:peak}
  \end{subfigure}  
  \caption{Numerical properties of the second (leading) maximum as a function of $\Lambda$. (a) The location of the new maximum $\delta^*$ exhibits a power-law decrease as a function of $\Lambda$. (b) The peak height $F(\delta^*,\Lambda)- F(0,\Lambda)$ decays exponentially as the coupling increases.}
  \label{fig:S}
\end{figure}

\newpage

\bibliography{references.bib}
\bibliographystyle{JHEP}

\end{document}